\documentclass[pre,aps,showpacs,preprintnumbers,amsmath,amssymb,superscriptaddress,twocolumn]{revtex4}
\usepackage{graphicx,subfigure}
\usepackage{pstricks}
\usepackage{xcolor}
\usepackage{enumerate}

\usepackage[normalem]{ulem}

\begin{document}

\title{Frequency-based brain networks: From a multiplex framework to a full multilayer description}

\author{J. M. Buld\'u}
\affiliation{Laboratory of Biological Networks, Center for Biomedical Technology (UPM), 28223, Pozuelo de Alarc\'{o}n, Madrid, Spain}
\affiliation{Complex Systems Group \& G.I.S.C., Universidad Rey Juan Carlos, 28933 M\'ostoles, Madrid, Spain}
\author{Mason A. Porter}
\affiliation{Department of Mathematics, University of California Los Angeles, Los Angeles, USA}
\affiliation{Oxford Centre for Industrial and Applied Mathematics, Mathematical Institute, University of Oxford, Oxford OX2 6GG, UK}
\affiliation {CABDyN Complexity Centre, University of Oxford, Oxford OX1 1HP, UK}
\begin{abstract}
We explore how to study dynamical interactions between brain regions using functional multilayer networks whose
layers represent different frequency bands at which a brain operates. Specifically, we investigate the consequences
of considering the brain \textcolor{black}{as (i) a multilayer network, in which all brain regions can interact with each other at different frequency bands; and (ii) as a multiplex network,} in which interactions between different frequency bands are allowed only within each brain region and not between them. We study the second-smallest eigenvalue $\lambda_2$ of the combinatorial supra-Laplacian matrix of both the multiplex and multilayer \textcolor{black}{networks, as $\lambda_2$ has been used previously as an indicator of network synchronizability and as a biomarker for several brain diseases.} We show that the heterogeneity of interlayer edges and, especially, the fraction of missing edges crucially modify the value of $\lambda_2$, and we illustrate our results with both synthetic network models and real data sets obtained from resting-state magnetoencephalography. \textcolor{black}{Our work highlights the differences between using a multiplex approach and a full multilayer approach when studying frequency-based multilayer brain networks}.
\end{abstract}


\maketitle


\section*{INTRODUCTION}

During the last few years, network science has undergone
a conceptual revolution from the extension of well-established techniques of network analysis
to multilayer networks \cite{kivela2014, boccaletti2014,dedomenico2013}, which provide a convenient way to simultaneously encode different types of interactions, subsystems, and other complications in networks. Consequently, it has been necessary to revisit our intuitive understanding of both structural and dynamical properties of networks --- including structural phase transitions \cite{radichi2013}, diffusion and other spreading processes \cite{dedom2016,salehi2015,gomez2013}, percolation and robustness \cite{buldyrev2010, gao2013}, synchronization \cite{aguirre2014}, and others --- to the new possibilities in multilayer descriptions, leading in many cases to counter-intuitive results.

The study of brain networks is currently undergoing a process of adaptation of classical single-layer (``monolayer'') concepts and analyses to a more general multilayer description. \textcolor{black}{(For reviews, see \cite{betzel2016,dedomenico2017,vaiana2017}; also see Fig.~1 of \cite{betzel2016} and Figs.~1--2 of \cite{dedomenico2017} for relevant schematics.)} Some studies have considered integration of data from structural and functional brain imaging into a multilayer network to account for both anatomical and dynamical information. In an early study using monolayer networks, \cite{honey2007} showed that a dynamical model simulated on the anatomical network of a macaque neocortex can successfully identify the positions of the anatomical hubs when signals are averaged appropriately. More recently, Stam et al. \cite{stam2016} analyzed how anatomical networks support activity, leading to specific functional 
networks (either undirected ones or directed ones), demonstrating that a dynamical model close to a critical transition is able to unveil interactions between structural and functional networks.

\textcolor{black}{It is also possible to combine anatomical and functional interactions of just a few nodes instead of an entire brain network.} For example, Battiston et al. \cite{battiston2016} examined network motifs (i.e., overrepresented network substructures that consist of a few nodes \cite{milo2002}) that combine anatomical connections (in one layer) and functional relations between cortical regions (in a second layer), linking data sets obtained, respectively, from diffusion-weighted magnetic resonance imaging (DW--MRI) and functional magnetic resonance imaging (fMRI). 

Efforts to combine anatomical and functional networks into a single multilayer network face the challenge of how to normalize the weights of edges that arise from different origins. To tackle this issue, Simas et al. \cite{simas2015} proposed translating functional and anatomical networks into a common embedding space and then comparing them in that space. They constructed functional networks (each with $N$ nodes) from the fMRIs of $q$ healthy individuals. They then used the functional networks of the $q=20$ individuals to construct a single functional multiplex network, a special type of multilayer network in which corresponding entities (brain regions) in different layers (individuals) can be connected to each other via interlayer edges, but other types of interlayer edges cannot occur \cite{kivela2014}. They followed a similar procedure to construct an anatomical multiplex network using data obtained from DW--MRI. Each of the multiplex networks was then independently projected into a common embedding space 
using a series of algebraic operations that allow one to calculate an ``algebraic aggregation'' of all layers into a single layer. (See \cite{simas2015} for details.) Using such a projection, it is possible to quantify the differences between anatomical and functional networks. Simas et al. also calculated an ``averaged aggregation'' of the functional and anatomical multiplex networks by 
averaging the weights of the corresponding edges over all layers.
They compared the two types of aggregation, and they were thereby able to identify certain brain regions (e.g., ones related to visual, auditory, and self-awareness processes) with significant differences between the functional and anatomical networks for both types of aggregation. However, only the algebraic aggregation was able to detect differences between the functional and anatomical networks in other regions (specifically, the thalamus, the amygdala, the postcentral gyrus, and the posterior cingulate), suggesting that the averaged aggregation disregards significant information \cite{simas2015}.

One possible alternative for reducing the complexity of analysis of brain networks is to concentrate only on ``functional'' (dynamical, in fact) interactions between brain regions and to define multilayer functional networks as a concatenation of a series of layers, each of which captures the interplay between brain regions during some time window. This approach, in which a layer in a multilayer network represents connection similarities over some time window, was taken in papers such as \cite{bassett2011,bassett2013,braun2014} to analyze the temporal evolution of network modules and examine dynamical reconfiguration and ``flexibility'' of functional networks.

Another alternative is to construct functional multilayer networks whose layers layers correspond to the well-known frequency bands at which a brain operates \cite{buzsaki2006}. As demonstrated by Brookes et al. \cite{brookes2016}, it is possible to construct {\it frequency-based multilayer networks} from magentoencephalographic (MEG) recordings by (i) band-pass filtering the raw MEG signals, (ii) obtaining the envelope of the amplitude at each frequency band, and (iii) measuring the pairwise correlations between each envelope (for whichever frequency it accounts). Using such an approach, \cite{brookes2016} constructed frequency-based multilayer networks, in which each layer includes the interactions in a given frequency band, and showed that the corresponding supra-adjacency matrices (which encode the linear-algebraic representation of connections in a multilayer network) convey statistically significant differences when comparing a control group with a group of schizophrenia sufferers. 

Recently, De Domenico et al. took the important step of analyzing the spectral properties of matrices associated with frequency-based multiplex networks \cite{dedomenico2016}. 
\textcolor{black}{The spectrum of an $\tilde{N} \times \tilde{N}$ matrix consists of the set $\{\lambda_1, \lambda_2, \lambda_3, \dots, \lambda_{\tilde{N}}\}$ of its eigenvalues, and it encodes valuable information about the structural properties of the corresponding network. In turn, these eigenvalues (as we will explain later) are related to various dynamical properties, such as network synchronizability, robustness, and diffusion \cite{newman2010}. In \cite{dedomenico2016}, De Domenico et al.} compared a group of schizophrenic patients with a control group using fMRI data, and they found that the second-smallest eigenvalue (i.e., the {\it algebraic connectivity} or ``Fiedler value'' \cite{miegh2011}) $\lambda_2$ of the combinatorial supra-Laplacian matrix \footnote{For such a matrix, $\lambda_1 = 0$.  \textcolor{black}{See the subsection ``Heterogeneity and missing interlayer edges in multiplex networks" for a detailed definition of the supra-Laplacian matrix.}} associated with a multiplex network is a better discriminator between the two groups than what one can obtain by studying the unfiltered or single-band functional networks (i.e., by using monolayer networks). De Domenico et al. also calculated centrality measures (i.e., measures of the importance of network components \cite{newman2010}) on the frequency-based multiplex networks to demonstrate the existence of hubs that had not been classified previously as important brain regions for functional integration. Hubs of the control group were located in anterior cingulate, superior frontal, insula, and superior temporal cortices; however, hubs for schizophrenic patients were distributed over frontal, parietal, and occipital cortices. These results revealed that frequency-based multiplex networks include relevant information about the functional organization of brain networks that is not captured by using a classical monolayer approach. 

\textcolor{black}{Multiple recent papers have demonstrated the benefits of using a multiplex description for analyzing the functional networks of patients suffering from Alzheimer's disease (AD). For example, Yu. et al. constructed frequency-based multiplex networks from MEG data and demonstrated that centrality measures calculated with layers analyzed
independently are unable to detect significant differences between a control group and AD group. However, when centralities are evaluated on a frequency-based multiplex network, one can find statistically significant differences in the hippocampus, posterior default-mode network, and occipital areas \cite{yu2017}.}
Guillon et al. also used frequency-based multiplex networks to differentiate between controls and people with AD \cite{guillon2017}. 
\textcolor{black}{In this case, the authors proposed the use of a multi-partitipation coefficient (MPC) to enhance classification of which individuals are suffering from AD. Their MPC
consists of an adaptation of the classical (i.e., monolayer) participation coefficient index \cite{guimera2005} to networks composed of several layers (see also \cite{yu2017}). A benefit
of using an MPC is that it does not depend on interlayer edge weights, which thus do not need to be calculated. As shown 
in \cite{guillon2017}, using a MPC increases classification accuracy and sensitivity when compared with monolayer network diagnostics.}

In this paper, we investigate how to translate the dynamics of different brain regions into a frequency-based (i.e., functional) multilayer network, in which individual layers account for coordination within a given frequency band. We focus specifically on the consequences of analyzing a multiplex network versus a more general multilayer one. The former allows interlayer connections only between the same brain region in different network layers, so coupling between oscillations in different frequency bands occurs only between the same brain region, whereas the latter allows one to model coordination between any brain region at any frequency band. We use resting-state MEG recordings because of their high temporal resolution (on the order of milliseconds), which makes it possible to analyze a broad spectrum of frequency bands \cite{depasquale2010,vandiessen2015}.
In our case, a set of MEG signals consists of $N$ time series, each of which comes from a sensor that captures the activity above a different cortical region. We then filter signals at four frequency bands (theta, alpha, beta, and gamma) and construct, \textcolor{black}{for each individual}, a 4-layer functional multilayer network from the dynamical coordination within and between frequency bands. 
\textcolor{black}{The existence of interlayer edges in frequency-based functional networks relies on the phenomenon of cross-frequency coupling (CFC), which is responsible of integrating brain activity at different spatial and temporal scales \cite{canolty2006}.  The quantification of CFC is a hard task, because the interplay between frequency bands is very intricate \cite{aru2015}. To address this issue, we set the weights of the interlayer edges to a value $p$, which one can estimate use an optimization function. 
For example, it is possible to construct multiplex networks of two groups of individuals with different profiles and determine a value of $p$ that best distinguishes between the two groups (see, e.g., \cite{dedomenico2016}). In our paper, we adopt a different strategy: we obtain the weights of the interlayer edges directly from time series using mutual information (MI) \cite{mackay}. We thereby capture the heterogeneity of both intralayer edges and interlayer edges, and we investigate the influence of such heterogeneity on spectral properties of frequency-based functional networks.}

Using both synthetic network models and data sets from laboratory experiments, we investigate 
the effects that heterogeneity of interlayer edges weights have on the spectral properties of both multiplex and more general multilayer
networks. Specifically, we focus on the algebraic connectivity $\lambda_2$, which is closely related to both structural and dynamical properties of networks \cite{newman2010,miegh2011,masuda2017}. On one hand, algebraic connectivity is an indicator of modular structure in networks \cite{fortunato2016}. In the framework of multilayer networks, one can interpret the value of $\lambda_2$ and how it changes as a function of interlayer coupling strength as a way to quantify structural integration and segregation of different network layers \cite{radichi2013} (also see the discussions in \cite{pauls2015}).  
On the other hand, $1/\lambda_2$ is proportional to the time required to reach equilibrium in a linear diffusion process \cite{gomez2013}. Additionally, the time $t_{\mathrm{sync}}$ to reach synchronization of an ensemble of phase oscillators that are linearly and diffusively coupled is also proportional to $1/\lambda_2$, and it is known that $t_{\mathrm{sync}}$ and $1/\lambda_2$ are positively correlated in some cases with nonlinear coupling \cite{almendral2007}.
\textcolor{black}{A recent investigation of the consequences of modifying interlayer edge weights of in multiplex networks have been has illustrated that enhancing the interlayer coupling shortens the transient regime to achieve interlayer synchronization in a Kuramoto model of coupled oscillators \cite{andrade2017}.}

\textcolor{black}{In the framework of functional brain networks, algebraic connectivity has been used as an indicator of Alzheimer Disease (AD), such as in \cite{dehaan2012}, who obtained statistically significant differences for $\lambda_2$ of functional networks obtained with MEG in a comparison of a group of patients suffering from AD with a control group. Phillips et al. 2015 \cite{phillips2015} calculated the value of $\lambda_2$ in a group of individuals with mild cognitive impairment and AD, although they did not report significant differences between them. Calculating $\lambda_2$ is also necessary for calculating the most standard type of synchronizability parameter, which is the ratio $\lambda_N/\lambda_2$, where $\lambda_N$ is the largest eigenvalue of the combinatorial Laplacian matrix \cite{boccaletti2006}. 
In a series of studies, the synchronizability of brain networks was calculated for different frequency bands \cite{basset2006}, during epileptic seizures \cite{schindler2008}, and for schizophrenic individuals \cite{siebenhuhner2013}.
We aim to improve the interpretation of algebraic connectivity for functional brain networks, and we thus investigate (1) how} the fact that a considerable fraction of all possible interlayer edges are not present in multiplex networks leads to a deviation from the theoretical values expected for $\lambda_2$ and (2) how these deviations are related to the mean weight of the interlayer edges. We thereby scrutinize the consequences of using a multiplex formalism, in which only cross-frequency coupling inside the same brain region is allowed, instead of employing a fully multilayer approach (i.e., without any restrictions to the type of coupling that one considers).

 \begin{figure}[!t]
\centering
\includegraphics[angle=0,width=0.50\textwidth]{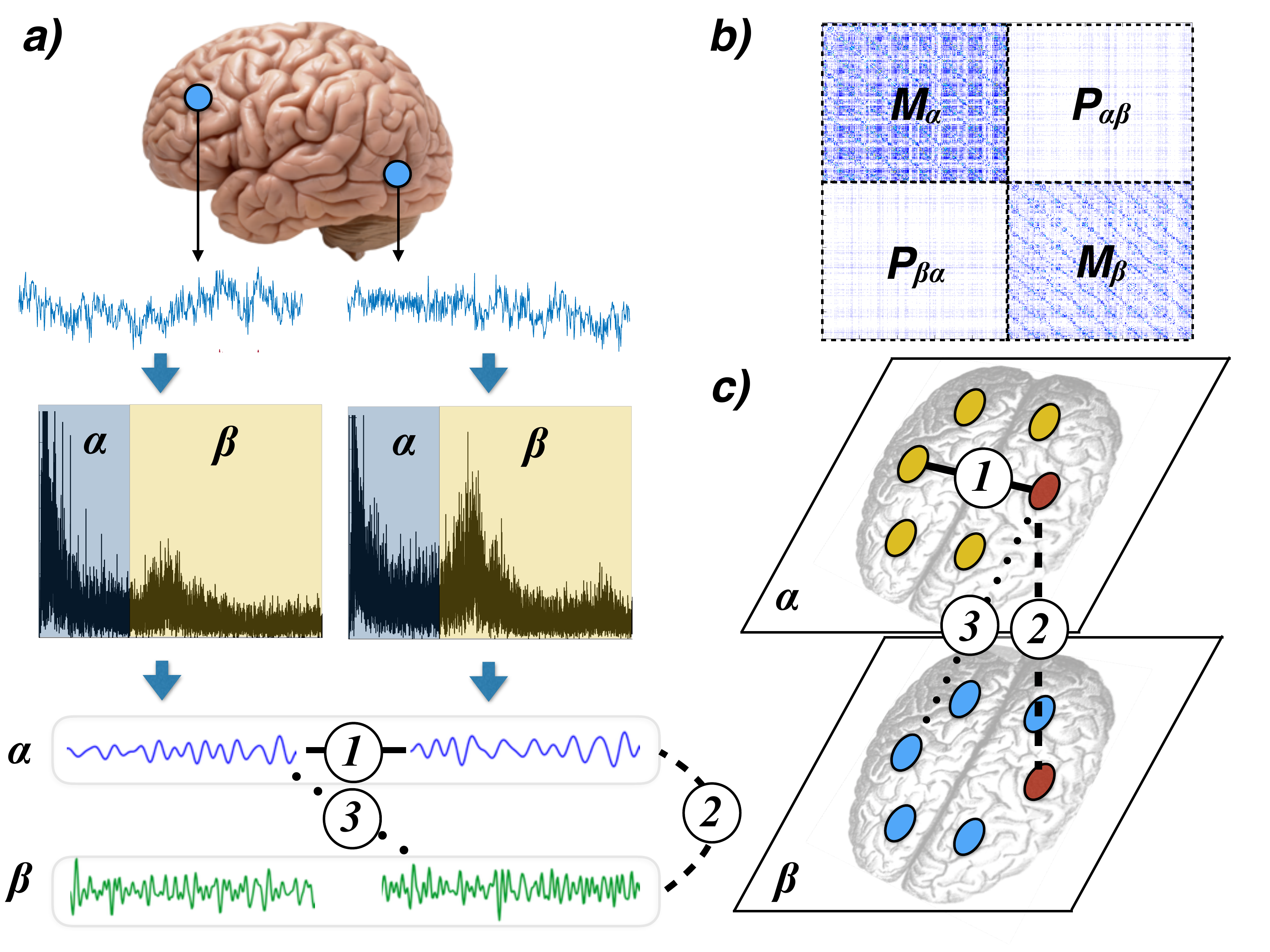}
\caption{{\bf Encoding brain dynamics as a multilayer functional network.} 
We show an illustrative example with two frequency bands (alpha and beta). (a) The MEG signals are band-pass filtered at two frequency bands: alpha [8--12] Hz and beta [12--30] Hz. We use mutual information (MI) \cite{mackay} to quantify the coordination between brain regions. This yields three different type of functional edges: edge ``1" quantifies the coordination between different regions at the same frequency band; edge ``2" corresponds to interlayer edges, which couple the activity of the same region at different frequency bands; and edge ``3" quantifies the cross-frequency coupling between different brain regions. Multiplex networks include only edges of types (1) and (2), whereas more general multilayer networks include all three types of edges. (b) Schematic of the supra-adjacency matrix of a 2-layer network constructed from the data in panel (a). (c) Schematic of the intralayer and interlayer edges in the multilayer functional network.
}
\label{fig:fig01}
\end{figure}

\section*{RESULTS}

\subsection*{Constructing frequency-based multilayer networks}

In Fig.~\ref{fig:fig01}, we illustrate the process of constructing frequency-based multiplex and multilayer brain networks.
Our starting point is a data set of MEG recordings of a group
of $q=89$ individuals during resting state (see Materials and Methods for details), but other experimental paradigms --- including different brain imaging techniques, such as fMRI or EEG --- can be used to construct multilayer networks with the same procedure.  
\textcolor{black}{Specifically, we record MEG activity \textcolor{black}{at $N$ cortical regions, with  \textcolor{black}{$235 \leq N \leq 246$} (see the Supplementary Information for details)}, and we then clean the data to remove artifacts and obtain corresponding 
unfiltered signals.} We thereby analyze the signal recorded by each sensor instead of carrying out a source reconstruction. 
We then band-pass filter each signal to obtain four different filtered time series for each brain region. We use the four classical frequency bands: theta [3--8] Hz, alpha [8--12] Hz, beta [12--30] Hz, and gamma [30--100] Hz. The number $l$ of layers of the multilayer network is the number of different frequencies that we examine (so $2 \leq l \leq 4$ in this case), and 
the $N$ nodes in each layer are associated with the dynamics of the $N$ sensors filtered at a given frequency band. We
number the nodes so that nodes $n$, $n+N$, $n+2N$, \dots,  $n+lN$ (with $n \in \{1, \dots N\}$) correspond to the signals of the same brain region $n$ at the $l$ different frequency bands (i.e., layers).

\begin{figure}[!t]
\centering
\includegraphics[angle=0,width=0.48\textwidth]{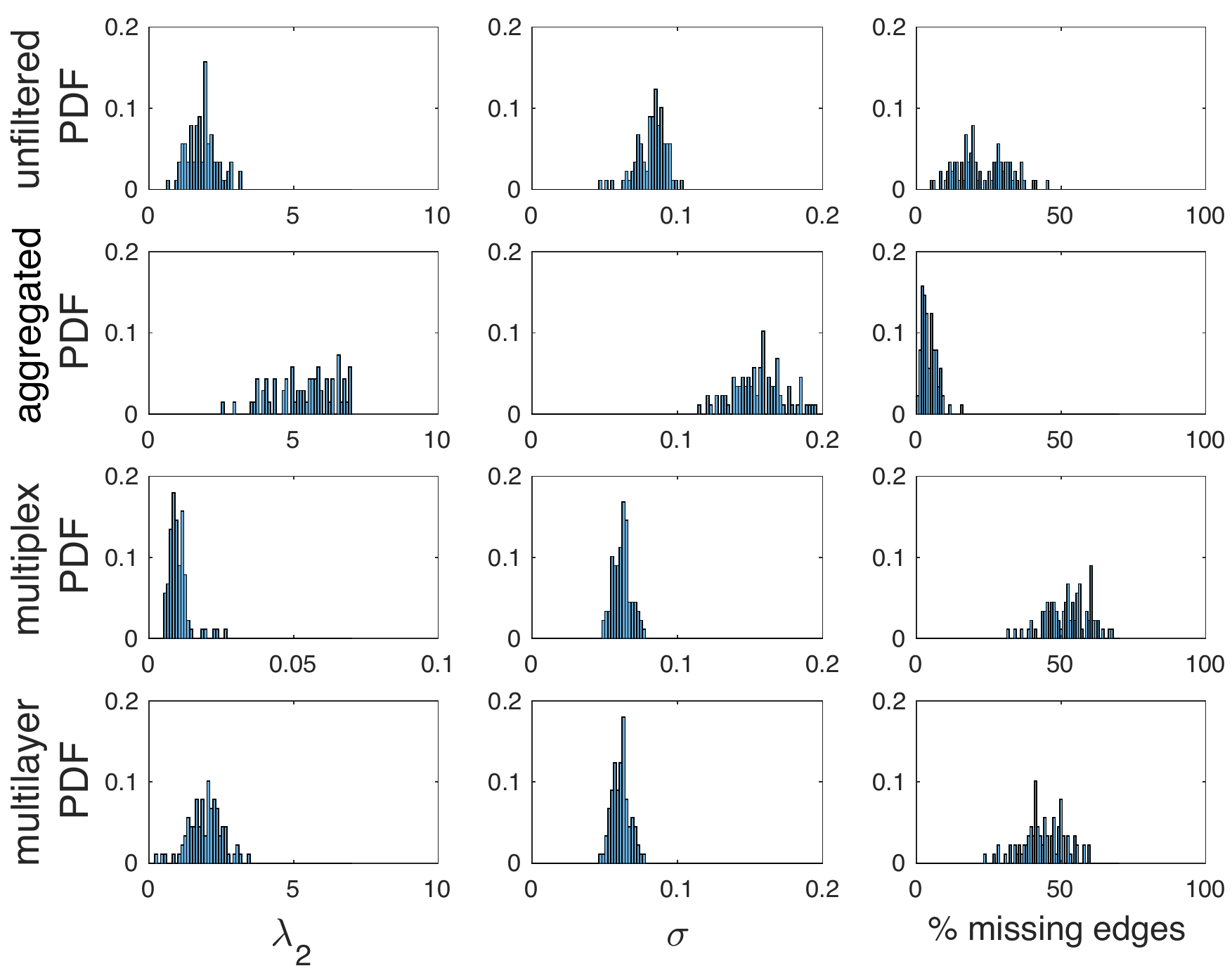}
\caption{{\bf MEG data sets: From unfiltered data to a multilayer network.} 
Probability distribution functions (PDFs) of different network characteristics of a group of $q=89$ individuals (see Materials and Methods for details).
We show the second-smallest eigenvalue $\lambda_2$ of the combinatorial Laplacian matrix, the standard deviation $\sigma$ of the matrix elements (to quantify their heterogeneity), and the percentage of missing edges of four different networks: (i) the functional network obtained from the unfiltered signals (first row), (ii) an aggregated network of the alpha and beta layers (second row), (iii) a multiplex network (third row), and a (iv) full multilayer network (fourth row).
In all cases, we only consider two layers (alpha and beta). \textcolor{black}{For the multiplex and multilayer networks, note that we are analyzing the supra-Laplacian matrices.}
The percentage of missing edges in the unfiltered and aggregated networks is equal to the percentage
of zeros in the whole matrix, but it refers only to the interlayer edges for the multiplex and multilayer networks.
}
\label{fig:fig02}
\end{figure}

We now quantify the coordination between any pair of nodes of a multilayer network, regardless of which layers they are in, using mutual information (see Materials and Methods). Calculating mutual information (MI) between time series of the same frequency band yields intralayer connections between brain regions (see edge ``1" in the bottom-left plot of Fig.~\ref{fig:fig01}a for an example), so each layer corresponds to a specific frequency band. Edges between the signals of the same sensor at different frequency bands result in interlayer connections between layers  (see edge ``2"). Such ``diagonal'' interlayer edges are the only type of interlayer edges that are allowed in multiplex networks \cite{kivela2014}.
Finally, cross-frequency coupling between different brain regions yield the other (``non-diagonal'') interlayer edges in a full multilayer network (see edge ``3"). As we show in Fig.~\ref{fig:fig01}(b), we thereby obtain a supra-adjacency matrix, where blocks along the diagonal account for intralayer connections (layers alpha and beta in the 2-layer example) and blocks off of the diagonal, marked as $\bf{P}^{\alpha\beta}$ and $\bf{P}^{\beta\alpha}$, encode the interlayer edges. Because $\mathrm{MI}_{ij}=\mathrm{MI}_{ji}$ for time series of nodes $i$ and $j$, the supra-adjacency matrix is symmetric, so ${\bf P}^{\alpha\beta}=[{\bf P}^{\beta\alpha}]^T$.

Importantly, although we have chosen to use MI, there are a diversity of similarity measures for capturing amplitude--amplitude and phase--amplitude correlations between different frequency bands (see \cite{aru2015} for a review on cross-frequency coupling measures), and each measure has its own advantages and drawbacks. Nevertheless, as we will see, the same methodological implications exist no matter which specific measure one uses to evaluate coordination between brain sites.

\begin{table}[!t] \centering
\begin{tabular}{|c|c|c|c|c|}
\hline \hline
 	Network					& $\bar\sigma$	& missing edges							\\ 
\hline \hline
unfiltered	 &  0.0823	 		& 	22.75\%	 \\ 
\hline
aggregated	 &  	0.1567 	&4.42\%	 \\ 
\hline
multiplex	 &  0.0618			&	52.17\%  \\
\hline
(full) multilayer	 & 0.0611		&	44.01\%	 \\ 
\hline \hline
\end{tabular}
\caption{For each kind of network (see the first column), we show the
mean standard deviation $\sigma$ of the weights of the edges (unfiltered and aggregated) and 
interlayer edges (for both multiplex networks and full multilayer networks) and the corresponding percentage of missing edges.
\textcolor{black}{(See the Supplementary Information for more details.)}
}
\label{TCexp}
\end{table}

In our discussion, we focus on the analysis of a 2-layer network with alpha and beta layers. We will discuss the consequences of considering alternative frequency bands and numbers of layers in the last section of the paper.

Our starting point is to compare the results from four different constructions of functional networks:
\begin{enumerate}[(i)]
\item  {\it Unfiltered functional
networks.} We obtain these networks from the original (unfiltered) signals of each brain region --- i.e., without decomposing the signals into different frequency bands --- so these are monolayer networks.
\item {\it Aggregated networks}. We obtain these networks from componentwise addition of
the weights of the alpha and beta layers to form monolayer networks.
\item {\it Multiplex networks}, in which each layer corresponds to a specific frequency band (as explained above) and interlayer edges are allowed only between nodes corresponding to the same brain region.
\item {\it Full multilayer networks}, which include the same layers as their multiplex counterparts, but with all possible interlayer edges.
\end{enumerate}

In Fig.~\ref{fig:fig02}, we show probability distribution functions (PDFs) for a group of $q=89$ individuals of the values of $\lambda_2$, 
the standard deviation of the interlayer edge weights (to quantify their heterogeneity), and the percentage of missing interlayer edges (also see Table \ref{TCexp}).

In column one of Fig.~\ref{fig:fig02}, we observe that the (monolayer) unfiltered functional network has similar a mean and standard deviation of $\lambda_2$ as the multilayer one (also see Table \ref{TCexp}). However, the aggregated networks tend to have larger
values of $\lambda_2$, which makes sense, as we construct an aggregated network by adding the weights of the two layers (alpha and beta), and the total ``strength'' of the network (i.e., the sum of all of its edge weights) is close to the double the strength of each layer. Importantly, the mean $\lambda_2$ for the multiplex networks is two orders of magnitude
smaller than the mean $\lambda_2$ of the multilayer networks. We expect this discrepancy, because we construct a multiplex network by deleting all interlayer edges of a full multilayer network, except for ones (so-called ``diagonal'' edges) that link the same ``physical" nodes. Thus, the total strength of the interlayer matrix $\bf{P}^{\alpha\beta}$ is considerably smaller in multiplex networks than in corresponding full multilayer networks. Because $\lambda_2$ is an indicator of the amount of interconnections between communities in a network \cite{newman2010}, one expects such drastic edge removals to yield a lower value of $\lambda_2$, as layers can be construed as communities with a small number of edges between them (only $N$ of the $N^2$ possible interlayer edges of the full multilayer network). 

In columns two and three of Fig.~\ref{fig:fig02}, we quantify the heterogeneity and the number of missing interlayer edges of the four different functional networks. In column two,
we plot the PDFs of the standard deviation of all edges (unfiltered and aggregated networks) and interlayer edges (multiplex and multilayer networks). In all cases, we observe that the functional networks obtained from the experimental data sets have non-negligible heterogeneity (also see Table \ref{TCexp}).
Again, aggregated networks have values that are roughly double those of the other kinds of networks because of the (rough) doubling of the mean strength.
In column three, we show \textcolor{black}{that there is a large percentage of missing edges. This arises from the fact that the edges whose weights are smaller than that obtained with an appropriate surrogate time series are not construed to be statistically significant, so we set their values to $0$.} (See Materials and Methods for details.)
The (monolayer) aggregated networks have the lowest mean percentage, followed by the (monolayer) unfiltered networks, and then the two types of multilayer networks. For the multiplex and full multilayer networks, the percentage of missing edges,
which is higher than 40\% in both cases, refers to the number of all possible interlayer edges.
Note that missing edges are unavoidable in functional brain networks, because not all brain regions communicate with each other through direct connections \footnote{In general, networks constructed from pairwise time-series similarities have nonzero edge weights in all (or almost all) intralayer adjacency entries \cite{bassett2017}. However, in functional brain networks, the deletion of entries that are not statistically significant can lead to a non-negligible number missing edges, as is the case with our data sets.}. Moreover, the different amounts of coordination between brain regions
lead to functional networks with heterogenous weights.

\subsection*{Heterogeneity and missing interlayer edges in multiplex networks}

Given our prior observations, a crucial question arises: What are the consequences of heterogeneity and missing interlayer edges, both intrinsic features of brain-imaging data sets, on multiplex and full multilayer functional networks? More specifically, how do they affect the value of $\lambda_2$ (and hence structural and dynamical implications that arise from such differences)?

With the aim of answering both questions, we do a series of numerical computations in which we compare the theoretical values of $\lambda_2$ in multiplex and full multilayer networks from homogeneous interlayer-edge distributions with ones from networks with heterogeneous and missing interlayer edges.

We start with a 2-layer multiplex network, whose layers alpha and beta have $N_\alpha =N_\beta$ nodes
and $L_\alpha=L_\beta$ intralayer edges, respectively. We number the nodes of the alpha layer from $k=1$ to $k=N_\alpha$ and the nodes of the beta layer \textcolor{black}{from $m=N_{\alpha}+1$ 
to $m = N_{\alpha\beta}=N_\alpha+N_\beta$. }
\textcolor{black}{The matrices ${\bf M^\alpha}$ and ${\bf M^\beta}$ are the corresponding adjacency matrices for each layer. We then introduce $l_c$ connector edges (i.e., diagonal interlayer edges) between each node $k$ of the alpha ayer to its corresponding node $m=k+N_{\alpha}$ of the beta layer to construct a multiplex network. We suppose that intralayer edges have weight $w^{\mathrm{intra}}_{i'j'}=1$  (i.e., for $i'$ and $j'$ belonging either to the alpha layer or to the beta layer), 
and we set the weights of the interlayer edges to $w^{\mathrm{inter}}_{km}=p_{km}$ (i.e., for $k  \in \alpha$ and $m \in \beta$), where $p_{km}$ are the elements of a vector $\vec{p}$ of the weights of the interlayer connections.}
Under these conditions, we obtain a supra-adjacency matrix ${\bf M^{\alpha\beta}}$ that consists of two diagonal blocks (${\bf M^\alpha}$ and ${\bf M^\beta}$) and two off-diagonal blocks (${\bf P^{\alpha\beta}}$ and ${\bf P^{\beta\alpha}}$, where ${\bf P^{\alpha\beta}}={\bf [P^{\beta\alpha}]^T}=\vec{p} \, \mathbb{I}$, where $\vec{p}$ is a row vector). That is,
\begin{equation}
	{\bf {M^{\alpha\beta}}} =
 \begin{pmatrix}
{\bf {M^\alpha}}  &  \vec{p} \, \mathbb{I}   \\
\vec{p} \, \mathbb{I}   & {\bf {M^\beta}} 
 \end{pmatrix}\,,
\end{equation}
where $\mathbb{I}$ is an identity matrix. \textcolor{black}{Following the same procedure, one can also extend the definition of a supra-adjacency matrix to an arbitrary number of layers. For example, if one considers layers for each
 of the theta, alpha, beta, and gamma bands (see Materials and Methods), one obtains the supra-adjacency matrix $M^{\theta\alpha\beta\gamma}$.}

The combinatorial supra-Laplacian matrix ${\bf \mathcal{L^{\alpha\beta}}}$ of the multiplex network is
\begin{equation}
	{\bf \mathcal{L^{\alpha\beta}} }=
 \begin{pmatrix}
{\bf  \mathcal{L}^{\alpha}} + \vec{p} \, \mathbb{I}  & - \vec{p} \, \mathbb{I}   \\
-\vec{p} \, \mathbb{I}   & {\bf \mathcal{L}^{\beta}} + \vec{p} \, \mathbb{I}
 \end{pmatrix}\,,
\end{equation}
where the layer combinatorial Laplacians $ \mathcal{L}^{\alpha,\beta}$ are
\begin{equation}
	{\bf \mathcal{L}}^{\alpha,\beta}_{ij}= \left\{
  \begin{array}{l l}
    s_i \,,& \quad \text{if $i=j$}\\
    -1 \,,& \quad \text{if $i$ and $j$ are adjacent}\\
		0 \,,& \quad \text{otherwise}\\
  \end{array} \right. \,,
\end{equation}
and $s_i=\sum_{i\neq j} w_{ij}$ is the weighted degree (i.e., total weight of incident edges) of node $i$.

\begin{figure}[!t]
\centering
\includegraphics[angle=0,width=0.4\textwidth]{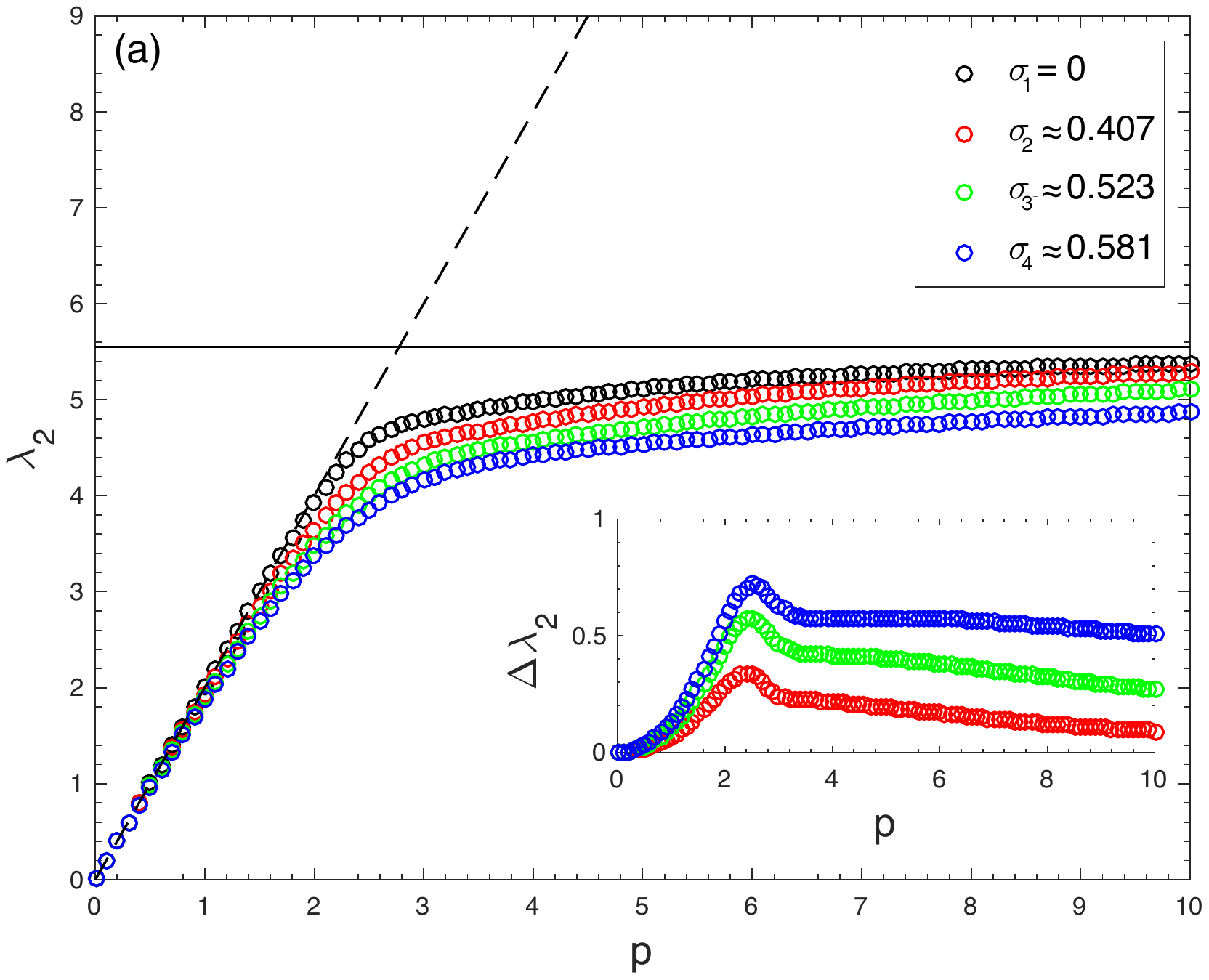} 
\includegraphics[angle=0,width=0.4\textwidth]{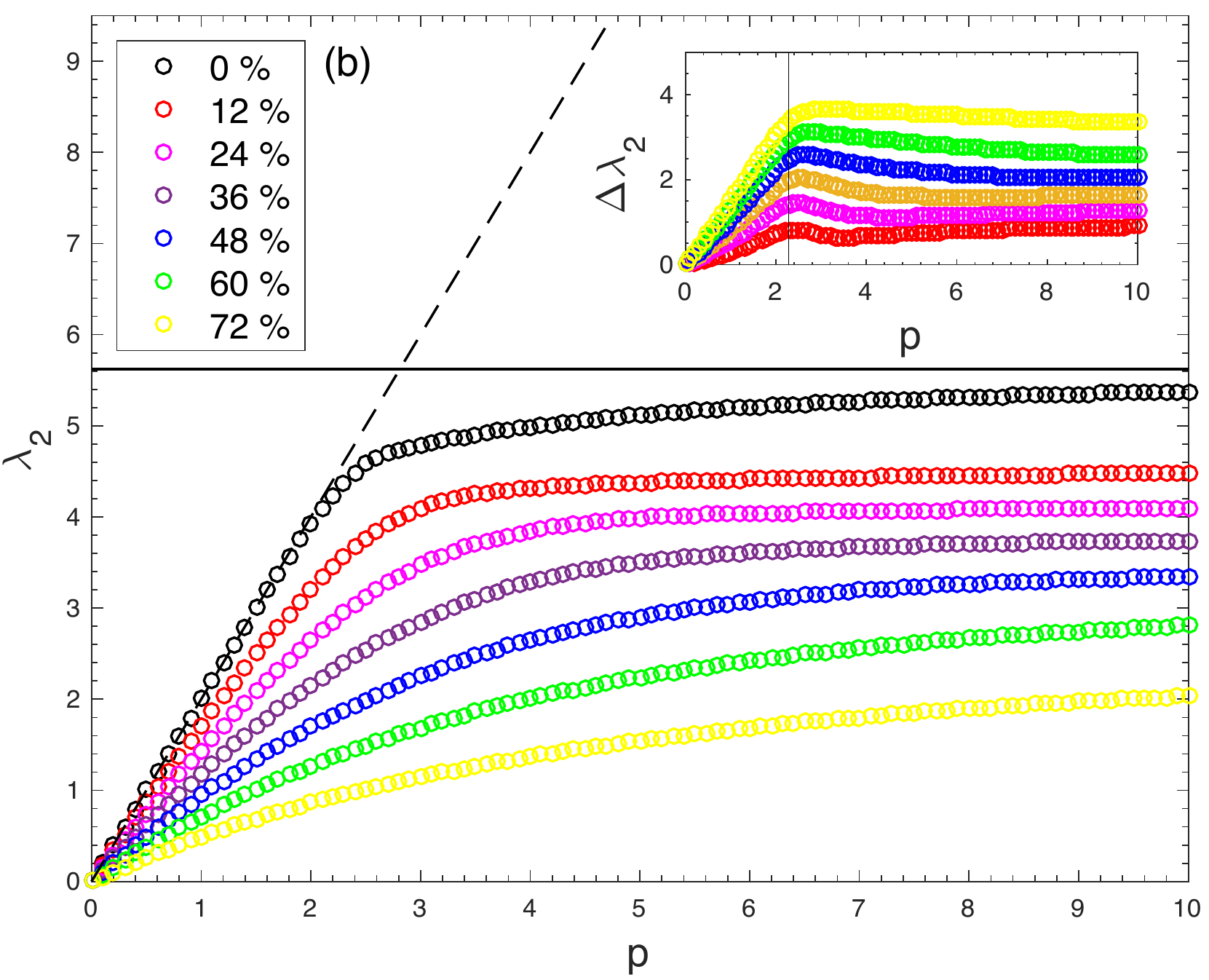}
\vskip-0.0cm
\caption{{\bf Consequences of heterogeneity and missing interlayer edges in a multiplex network.} 
Algebraic connectivity $\lambda_2$ of the combinatorial supra-Laplacian matrix $\mathcal{L}^{\alpha \beta}$. Each of the two layers is a $G(N,p_{\mathrm{con}})$ Erd\H{o}s--R\'enyi network with $N=250$ nodes and a connection probability of 
$p_{\mathrm{con}}=0.25$. Each circle corresponds to a mean over 100 realizations.
{\bf (a)} We quantify the heterogeneity of the interlayer edges with the standard deviation $\sigma$ of their weights: (i) $\sigma=0$ (black circles), (ii) $\sigma \approx 0.407$ (red circles), (iii) $\sigma \approx 0.523$ (green circles), and (iv) $\sigma \approx 0.581$ (blue circles). 
 Lines correspond to analytical solutions for the case $\sigma=0$ (i.e., a homogenous multiplex network). The dashed line is $\lambda_2= 2p$, and the solid line is $\lambda_{2,\mathrm{agg}}= \frac{1}{2}\lambda_2(\mathcal{L}^\alpha + \mathcal{L}^\beta)$
\cite{radichi2013,sahneh2015}. In the inset, we show $\Delta \lambda_2$, the difference of $\lambda_2$ between the homogeneous multiplex network and multiplex networks with heterogeneous interlayer edges. {\bf (b)} Algebraic connectivity $\lambda_2$ of the multiplex network as a function of the number of missing interlayer edges. In the inset, we show $\Delta \lambda_2$, the difference of $\lambda_2$ between the homogeneous multiplex network (i.e., $\sigma=0$ and all possible interlayer edges) and the multiplex networks with missing interlayer edges. The solid vertical line indicates the mean value $\langle p^* \rangle$ of the transition point.
}
\label{fig:fig03}
\end{figure}

In Fig.~\ref{fig:fig03}, we show the consequences of heterogeneity on the distribution of the weights of interlayer edges of the multiplex network ${\bf M^{\alpha\beta}}$.
In this example, the multiplex network has an interlayer connection matrix ${\bf P^{\alpha\beta}}= \vec{p} \, \mathbb{I}$, where $\mathbb{I}$ is an $N_\alpha \times N_\alpha$ (equivalently, $N_\beta \times N_\beta$, as $N_\alpha = N_\beta$ in this example) identity matrix and $\vec{p}=p \vec{h}$ is a row vector controlling the weights of interlayer edges, where $p$ modulates its amplitude and the vector $\vec{h}$ encodes the heterogeneity of the interlayer edges. We set the elements of $\vec{h}$ to follow a uniform distribution over the interval [$h_{\mathrm{min}}, h_{\mathrm{max}} ]$. These elements have a mean value of
$\bar{h}$ and a standard deviation of $\sigma$. We set $\bar{h}=1$ and construct networks with interlayer-edge-weight heterogeneities that range from $\sigma=0$ (blue circles) to $\sigma \approx 0.581$ (green circles). We then analyze the interplay between the weights of the interlayer edges and the heterogeneity by increasing the value of $p$. Note that $\sigma=0$ corresponds to what we call a {\it homogeneous multiplex network}, which has uniformly-weighted interlayer edges (i.e., $\vec{p}=p$ for all interlayer edges). We obtain the results in Fig.~\ref{fig:fig03} from a mean over 100 realizations of 2-layer networks with the $G(N,p_{\mathrm{con}}$ Erd\H{o}s--R\'enyi (ER) model (with $p_{\mathrm{con}}=0.25$) in each layer and $N = N_\alpha=N_\beta=250$ nodes \cite{newman2010}.

As explained in \cite{radichi2013}, modifying the weight parameter $p$ of the interlayer edges has important consequences for the value of $\lambda_2$ for homogeneous multiplex networks. 
The existence of two regimes of qualitatively distinct dynamics, separated at a transition point $p^*$, was discussed in \cite{radichi2013} (and in various subsequent papers): when $p\ll p^*$, the algebraic connectivity $\lambda_2$ follows the linear relation $\lambda_2=2 p$; for $p \gg p^*$, however, the value of $\lambda_2$ approximates that of the aggregated network [i.e., $\lambda_{2,\mathrm{agg}}= \frac{1}{2}  \lambda_2(\mathcal{L}^\alpha + \mathcal{L}^\beta)$].
In Fig.~\ref{fig:fig03}, the dashed and solid lines indicate the theoretical predictions of $\lambda_2$ in the homogeneous case for small (dashed) and large (solid) values of $p$. We obtain a mean value of $p^* \approx 2.870$ by locating the intersection between the eigenvalues $\lambda_2$ and $\lambda_3$ of the supra-Laplacian matrix $\mathcal{L}^{\alpha \beta}$ \cite{sahneh2015}.

Figure~\ref{fig:fig03}(a) illustrates that the heterogeneity of the interlayer edges leads to non-negligible differences in
$\lambda_2$. Furthermore, values of $p$ near $p^*$ have the maximum discrepancy between the heterogeneous case (colored circles) and the homogeneous case (black circles). \textcolor{black}{In studies of multiplex networks, it is simplest to take interlayer edge weights to be homogeneous (i.e., given by a constant value $p$), especially when it is not clear how to estimate their values \cite{kivela2014}. When possible, however, it is desirable to be more sophisticated, and one can quantify the coupling strength between different frequencies in various ways when studying frequency-based functional brain networks \cite{aru2015}. The choice of incorporating versus ignoring heterogeneity of inter-frequency edge weights then leads to unavoidable differences in the estimation of $\lambda_2$, especially near the transition point $p^*$.}
  
\textcolor{black}{These discrepancies in the value} of $\lambda_2$ are even larger
when some interlayer edges are missing.
 In Fig.~\ref{fig:fig03}(b), we show the effects of removing some percentage of the interlayer edges in a ``holey" multiplex network (i.e., a multiplex network with an interlayer matrix ${\bf P^{\alpha\beta}}$ whose elements are either $1$ or $0$ in the diagonal and $0$ in all other elements). We observe that increasing the number of missing interlayer edges causes the multiplex networks to have drastically reduced values of $\lambda_2$. When  $p \ll p^*$, the value of $\lambda_2$ always grows with a slope that is smaller than $2p$; for 
$p \gg p^*$, however, the value of $\lambda_2$ never reaches the value of $\lambda_2$ for the aggregated network. Finally, values of $p$ close to $p^*$ again have the largest discrepancies with respect to the homogeneous case.

\section*{Mutilplex networks versus full multilayer networks}
\begin{figure}[!t]
\centering
\includegraphics[angle=0,width=0.4\textwidth]{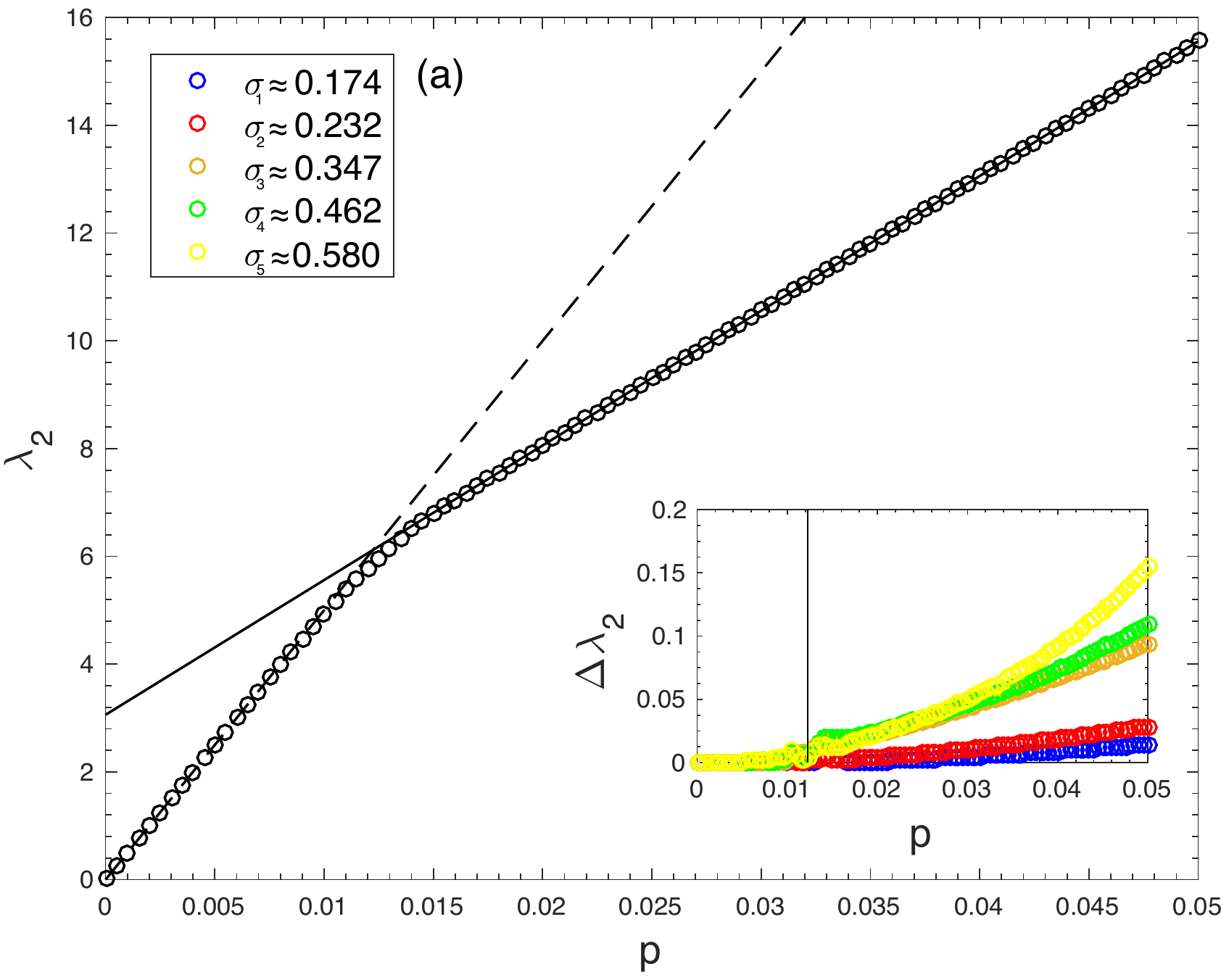}
\includegraphics[angle=0,width=0.4\textwidth]{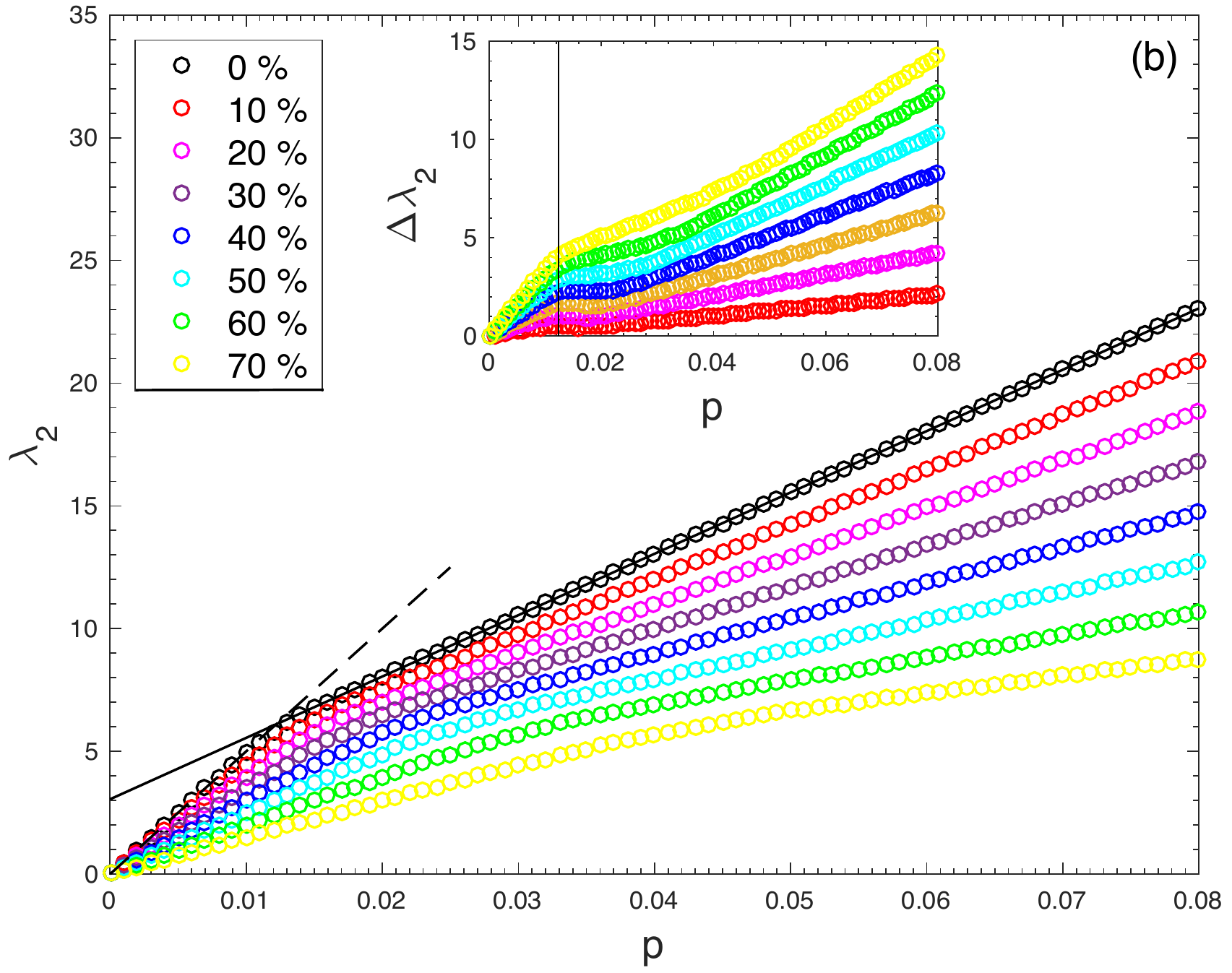}
\vskip-0.cm
\caption{{\bf Edge heterogeneity and missing interlayer edges in a full multilayer network.} 
Algebraic connectivity $\lambda_2$ of the combinatorial supra-Laplacian matrix $\mathcal{L}_{\alpha \beta}$ (black circles) as a function of the weight parameter $p$ of the interlayer connections. Each of the two layers is a $G(N,p_{\mathrm{con}})$ Erd\H{o}s--R\'enyi network with $N=250$ nodes and a connection probability of $p_{\mathrm{con}}=0.25$. Each circles corresponds to a mean over 100 realizations of such multilayer networks.  {\bf (a)} Algebraic connectivity $\lambda_2$ for a homogeneous interlayer connectivity matrix $\bf{P}^{\alpha\beta}$=$p\bf{C}$, where each of the elements of $\bf{C}$ is $c_{ij}=1$. 
The dashed line, given by $\lambda_2=2p \langle c \rangle$, corresponds to the analytical solution for a homogeneous matrix $\bf{P}^{\alpha\beta}$, where $\bf{C}$ has a mean weighted degree of $\langle c \rangle=250$. 
The solid line is given by $\lambda_2=\lambda_{2,\mathrm{min}(\alpha,\beta)}+\lambda_2(p\bf{C})$, 
where $\lambda_{2,\mathrm{min}(\alpha,\beta)}$ is the value of the smaller $\lambda_2$
of the two isolated layers. The inset shows the differences $\Delta \lambda_2$ between a multilayer network with homogeneous interlayer connectivity matrix $\bf{P}^{\alpha\beta}$ and a series of multilayer networks
with an increasing heterogeneity (quantified by the standard deviation $\sigma$) in $\bf{P}^{\alpha\beta}$. Differences with the homogeneous case increase after the mean of the transition point $\langle p^* \rangle$ (vertical solid line). {\bf (b)}
Algebraic connectivity $\lambda_2$ of the full multilayer network as a function of the percentage of missing interlayer edges (see the figure legend). In the inset, we show the increment of $\lambda_2$ versus the number of missing interlayer edges.
}
\label{fig:fig04}
\end{figure}

We now identify the qualitative differences (and some of their consequences) between multiplex networks and full multilayer networks. The latter have interlayer connectivity matrix ${\bf P^{\alpha\beta}}=p{\bf C}$, where $\bf{C}$ has elements $c_{ij}$ that account for the weight between each pair $\{i,j\}$ of nodes $i \in \alpha$ and $j \in \beta$, and the parameter $p$ allows one to modulate the mean weight of the interlayer edges. In Fig.~\ref{fig:fig04}(a), we connect the same layers as in the previous section, but now we use a homogeneous interlayer connectivity matrix ${\bf P^{\alpha\beta}}$ with weights $p_{ij}=p$ (i.e., $c_{ij}=1$ for all $i$ and $j$). Increasing $p$ from $0$ leads to the transition point $p^* \approx 0.013$, which we obtain from the intersection of the eigenvalues $\lambda_2$ and $\lambda_3$ of the combinatorial supra-Laplacian $\mathcal{L^{\alpha \beta}}$. We observe for $p<p^*$ that the value of $\lambda_2$ follows the linear function $\lambda_2= 2 p \langle c \rangle$ [the dashed line in Fig.~\ref{fig:fig04}(a)], where $\langle c \rangle=250$ is the mean weighted degree of the matrix $\bf{C}$, as demonstrated in \cite{radichi2013}. 
Interestingly, after the transition point $p = p^*$, the value of $\lambda_2$ can be described by the function $\lambda_2=\lambda_{2,\mathrm{min}(\alpha,\beta)}+\lambda_2(p{\bf C})$, where $\lambda_{2,\mathrm{min}(\alpha,\beta)}$ is the value of the smaller $\lambda_2$ of the two isolated layers. As we show in the inset of Fig.~\ref{fig:fig04}(a), when a certain amount of heterogeneity is introduced into the interlayer connectivity matrix, we observe \textcolor{black}{slight differences from the homogeneous case. Specifically, for the heterogeneous case, they are larger} for values of $p$ above \textcolor{black}{the mean $\langle p^* \rangle$ of the transition point.}

As we observed in the analogous scenario for multiplex networks, deleting interlayer edges of a full multilayer network increases the differences of $\lambda_{2}$ compared to that of a homogeneous multilayer network. 
As we show in Fig.~\ref{fig:fig04}(b), the value of $\lambda_2$ decreases with the number of missing edges, which one should expect, because having a smaller number of interlayer edges implies that one needs larger connection weights to maintain the same amount of interlayer coupling. Interestingly, the deviation from the theoretical predictions
is significant even for $p < \langle p^* \rangle$. Additionally, for $p>\langle p^* \rangle$, we observe the expected change in the slope of $\lambda_2$, but the theoretical predictions given by $\lambda_2=\lambda_{2,\mathrm{min}(\alpha,\beta)}+\lambda_2(p\bf{C})$ (solid line) begin to fail, leading to a discrepancy that increases with the number of missing edges.

\begin{figure}[!t]
\centering
\hskip-0.5cm
\includegraphics[angle=0,width=0.48\textwidth]{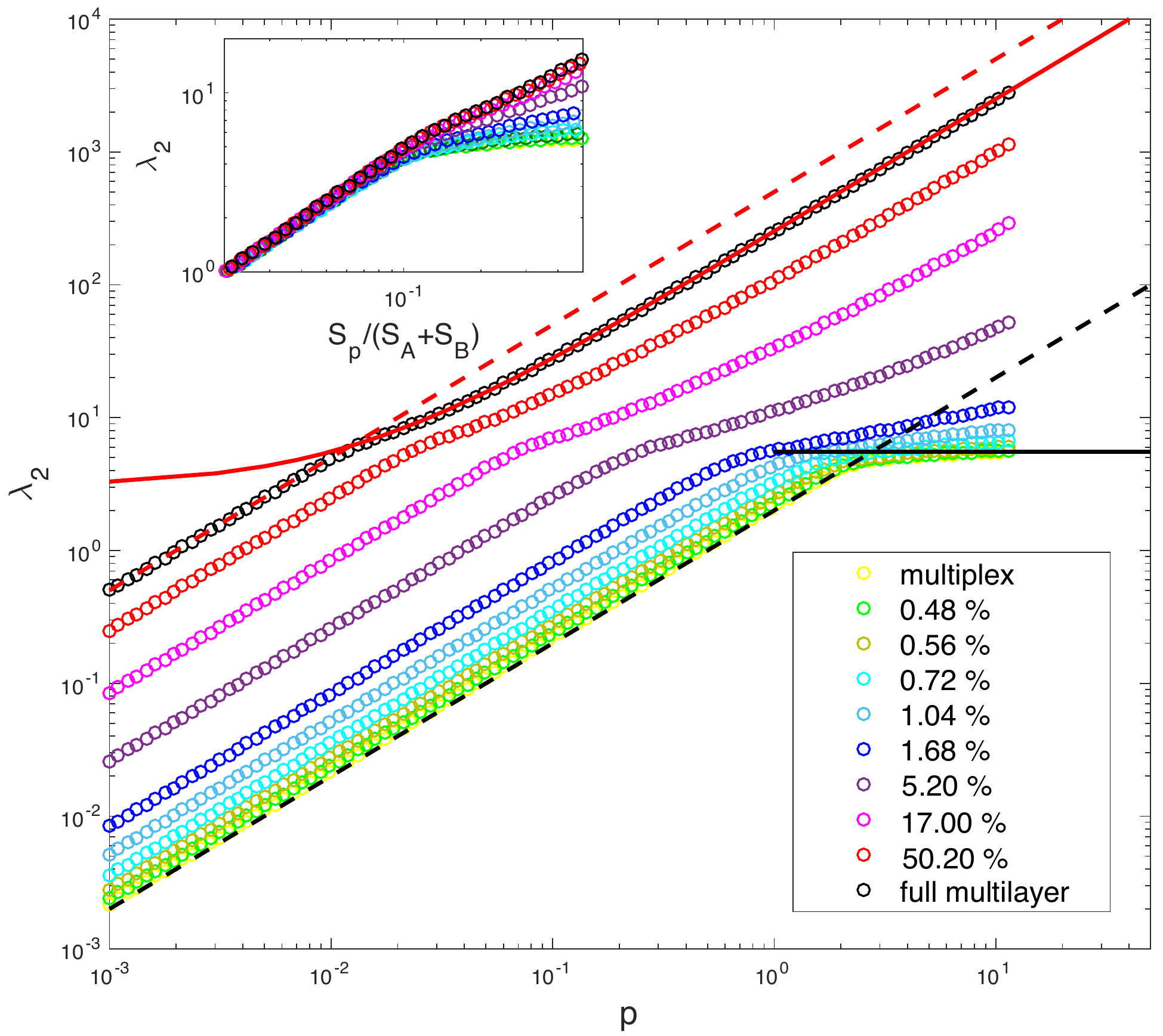}
\vskip-0.cm
\caption{{\bf Transition from a multiplex network to a full multilayer network.} 
Algebraic connectivity $\lambda_2$ of the combinatorial supra-Laplacian matrix $\mathcal{L}^{\alpha \beta}$ for different values of the percentage of edges in the interlayer connectivity matrix $\bf{P}^{\alpha\beta}$. Departing from a multiplex network, we add interlayer edges uniformly at random and calculate the percentage of existing
edges in $\bf{P}^{\alpha\beta}$. Each circle corresponds to a mean over 100 realizations. We set all active edges of the interlayer connectivity matrix $\bf{P}^{\alpha\beta}$ to $p_{ij}=p$. The dashed lines correspond to the analytical solutions for the full multilayer network (black dashed line; $\lambda_2=2p \langle c \rangle$, with $\langle c \rangle=250$) and the multiplex network (red dashed line; $\lambda_2=2p$) for $p<\langle p^* \rangle$. The solid curves are the theoretical solutions when $p>\langle p^* \rangle$ for the full multilayer network (black line; $\lambda_2=\lambda_{2,\mathrm{min}(\alpha,\beta)}+\lambda_2(p\bf{C})$) and the multiplex network (red curve; $\lambda_{2,\mathrm{agg}}= \frac{1}{2} \lambda_2(\mathcal{L}^A + \mathcal{L}^B)$). In the inset, we show $\lambda_2$ versus $S_P/(S_\alpha + S_\beta)$, where $S_P$, $S_\alpha$, and $S_\beta$ are, respectively, the strength of the interlayer matrix $\bf{P}^{\alpha\beta}$, the strength of the alpha layer, and the strength of the beta layer. Note that differences between the values of $\lambda_2$ of the multilayer structures increase significantly for $S_P/(S_\alpha + S_\beta) \gtrapprox 0.1$.
}
\label{fig:fig05}
\end{figure}

Because full multilayer networks have up to $N^2$ interlayer edges, whereas multiplex networks can have only $N$ of them, the former tend to have interlayer connectivity matrices with higher strengths $S_P=\sum_{ij} p_{ij}$.
In Fig.~\ref{fig:fig05}, we show the algebraic connectivity $\lambda_2$ of the combinatorial supra-Laplacian matrix $\mathcal{L}^{\alpha \beta}$ for a series of networks with identical layers and interlayer strength $S_P$, but with a different number of interlayer edges, ranging from a multiplex network ($N$ interlayer edges) to a full multilayer network with no nonzero entries ($N^2$ interlayer edges). The dashed lines correspond to the analytical solutions for the full multilayer network (black dashed line; $\lambda_2=2p \langle c \rangle$, with $\langle c \rangle=250$) and the multiplex network (red dashed line; $\lambda_2=2p$) when $p<p^*$. The solid curves are the corresponding theoretical solutions when $p>\langle p^* \rangle$ for the full multilayer network [black line; $\lambda_2=\lambda_{2,\mathrm{min}(\alpha,\beta)}+\lambda_2(p\bf{C})$] and the multiplex network [red curve; $\lambda_{2,\mathrm{agg}}= \frac{1}{2} \lambda_2(\mathcal{L}^A + \mathcal{L}^B)$]. We observe the effect that adding interlayer edges to multiplex networks has on the value of $\lambda_2$ and the associated transition from a multiplex network to a full multilayer architecture. Interestingly, the different numbers of interlayer edges in the two types of networks leads to a difference in the position of $\langle p^* \rangle$ (which can be inferred by looking at the change of slope of $\lambda_2$) that can reach several orders of magnitude.

The inset of Fig.~\ref{fig:fig05} illustrates the same results normalized by the total strength of the interlayer connectivity matrices. That is, we show $S_P/(S_\alpha + S_\beta)$, where $S_\alpha=\sum M^{\alpha}_{ij}$ and $S_\beta=\sum M^{\beta}_{ij}$, respectively, are the strengths of layers alpha and beta. This allows us to compare networks with the same value of $S_P$, regardless of whether they are close to a multiplex architecture or to a fully multilayer architecture with no nonzero entries. Below $\langle p^* \rangle$ we obtain similar values of $\lambda_2$ for all network architectures. It is only for $p>\langle p^* \rangle$ that the particular structure of the interlayer connectivity matrix begins to play a role in the value of $\lambda_2$. We observe that differences start to arise at $S_P/(S_\alpha + S_\beta) \approx 0.1$, which is a relatively low value.

Although we used ER intralayer networks in our above calculations, we obtain similar results for other network models. In particular, the results are qualitatively the same when we construct the intralayer networks using a Barab\'asi--Albert (BA) model \cite{barabasi1999}. See the Supplementary Information for details.

\section*{The meaning of $\lambda_2$ in experimental data}

Now that we have analyzed the effects of weight heterogeneity and the number of missing interlayer edges, let's revisit the multiplex and 
multilayer networks obtained from the MEG recordings. As we saw in Fig.~\ref{fig:fig02}, both edge-weight heterogeneity and missing interlayer edges occur in our experimental data, and it is thus desirable to investigate how close our networks constructed from experimental data are to a transition point $p^*$ and how this proximity (or lack thereof) influences the expected value of $\lambda_2$.

\begin{figure}[!t]
\centering
\includegraphics[angle=0,width=0.50\textwidth]{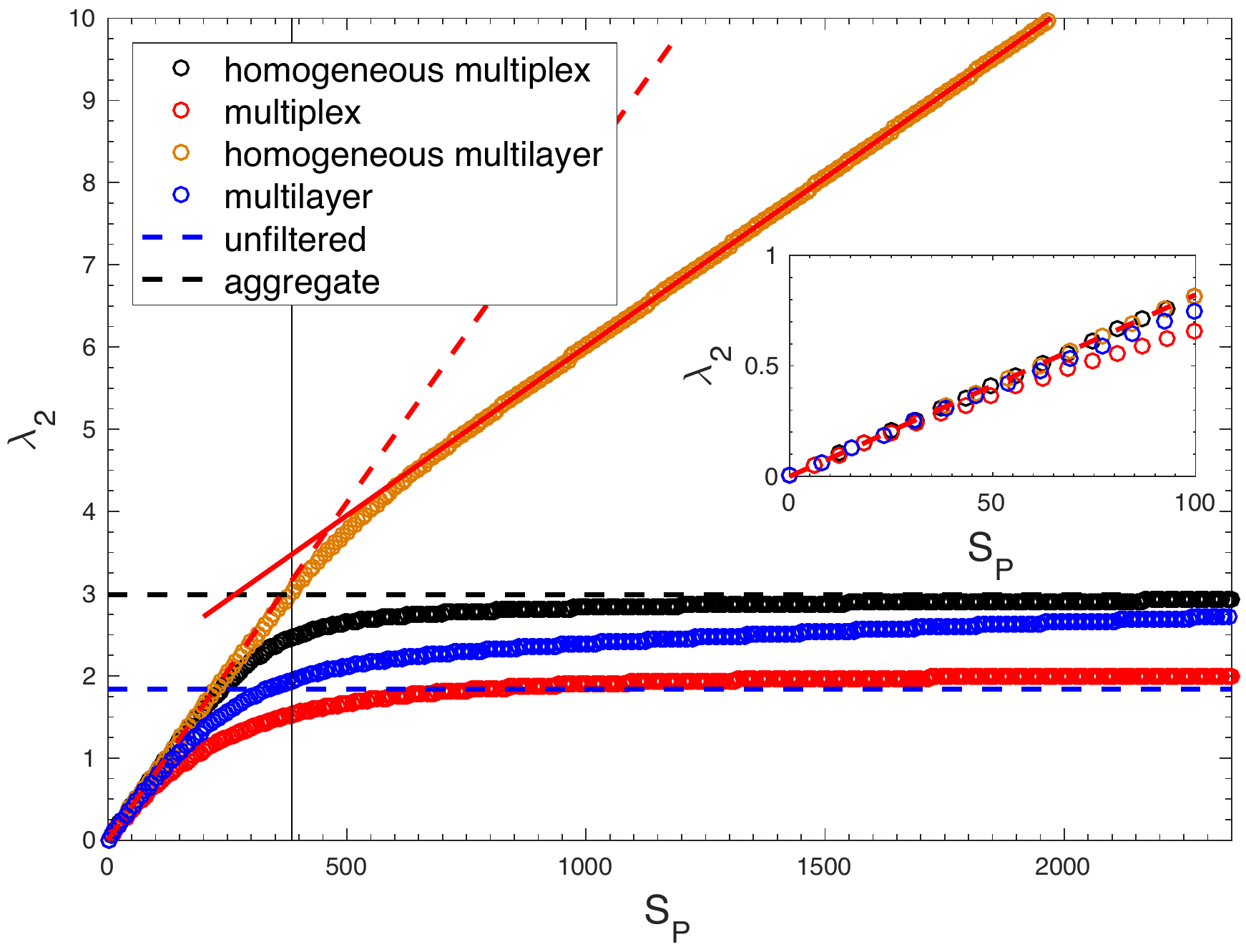}
\caption{{\bf Edge heterogeneity and missing interlayer edges in frequency-based functional brain networks.} 
Algebraic connectivity $\lambda_2$ of the combinatorial supra-Laplacian matrix $\mathcal{L}^{\alpha \beta}$ as a function of the mean strength of the interlayer connectivity matrix $\bf{P}^{\alpha\beta}$$=p\bf{C}$, where $\bf{C}$ encodes the weights of the interlayer edges, for four different types of networks: (i) homogeneous multiplex networks (black), (ii) heterogeneous multiplex networks (red), (iii) homogeneous full multilayer networks (orange), and (iv) heterogeneous full multilayer networks (blue). 
Each circle is a mean over $89$ individuals. The blue and black dashed lines, respectively, are the values of $\lambda_2$ for the unfiltered and aggregated networks. The red dashed line corresponds to $\lambda_2=2p \langle c \rangle$, where $ \langle c \rangle$ is the mean of the weighted degree. 
The red solid line is given by $\lambda_2=\lambda_{2,\mathrm{min}(\alpha,\beta)}+\lambda_2(p\bf{C})$, 
where $\lambda_{2,\mathrm{min}(\alpha,\beta)}$ is the value of the smaller $\lambda_2$
of the two isolated layers. 
}
\label{fig:fig06}
\end{figure}

In Fig.~\ref{fig:fig06}, we show the values of $\lambda_2$ that we obtain for four different network reconstructions based on the MEG data: a homogeneous multiplex network (black circles), a heterogeneous multiplex network (red circles), a homogeneous full multilayer network (blue circles), and a heterogeneous full multilayer network (cyan circles).
Each network has two layers --- one for the alpha band and one for the beta band --- and each point corresponds to the mean over the group of $89$ individuals. 
We obtain the heterogeneous versions of the multiplex and multilayer networks using the MI values between the brain regions in frequency bands, as we described previously (also see Materials and Methods). Second, we construct the homogeneous versions of both the multiplex and multilayer networks by assigning the same weight $\langle{c}\rangle$ to all interlayer edges, where $\langle{c}\rangle$ is the mean of the weights of the interlayer edges in their heterogeneous counterparts. Note
that, in this case, the homogeneous multiplex and multilayer networks do not correspond to real functional networks, but instead they are reference networks that we use to quantify the consequences on $\lambda_2$ of the intrinsic heterogeneity and missing edges in real functional networks.

To assess how close the real networks are to the transition point $p^*$, we multiply the value of the interlayer edges by a parameter $p$, which we increase from $p=0$ to $p>1$. We then calculate the strength $S_{P}$ of the corresponding interlayer connectivity matrix $\bf{P^{\alpha\beta}}$ to allow a comparison between the multiplex network and the multilayer networks, and we then plot the value of $\lambda_2$ versus $S_{P} $. The red dashed line in Fig.~\ref{fig:fig06} corresponds to the theoretical predictions for $p<p^*$ (i.e., $\lambda_2=2p\langle c \rangle$, where  $\langle c \rangle$ is the mean weighted degree of the nodes in the interlayer connection matrix $\bf{P^{\alpha\beta}}$). 
The black and blue dashed lines are, respectively, the value of $\lambda_2$ for the aggregated network divided by $2$ [i.e., $\lambda_{2,\mathrm{agg}}= \frac{1}{2} \lambda_2(\mathcal{L}^\alpha + \mathcal{L}^\beta)$] and the value of $\lambda_2$ for the unfiltered (monolayer) functional network.
The vertical solid line corresponds to the case $p=1$ for the mean of the heterogeneous full multilayers networks (i.e., the networks that we obtain by taking into account all interlayer correlations without modifying their weights). Interestingly, this network occupies the region in which the evolution of $\lambda_2$ with respect to $p$ is changing slope from $\lambda_2=2p \langle c \rangle$ to $\lambda_2=\lambda_{2,\mathrm{min}(\alpha,\beta)}+\lambda_2(p\bf{C})$, which suggests that the frequency-based multilayer networks that we analyze are close to the transition point $p^*$. 
As we have seen, it is near this point where the value of $\lambda_2$ is most influenced by the effects of heterogeneity and missing interlayer edges. 
\textcolor{black}{It is also worth noting that the four network representations have rather different values of algebraic connectivity (and hence, we expect, of spectral properties more generally), except when $p\ll p^*$ (see the inset of Fig.~\ref{fig:fig06}).
Although it looks from the plot that the Fiedler values of the heterogeneous multilayer networks (cyan circles) and those of the homogeneous multiplex networks (black circles) may converge to the same value for large $S_P$, they eventually cross each other when $S_P$ is further increased (not shown).}


\begin{figure}[!ht]
\centering
\includegraphics[angle=0,width=0.45\textwidth]{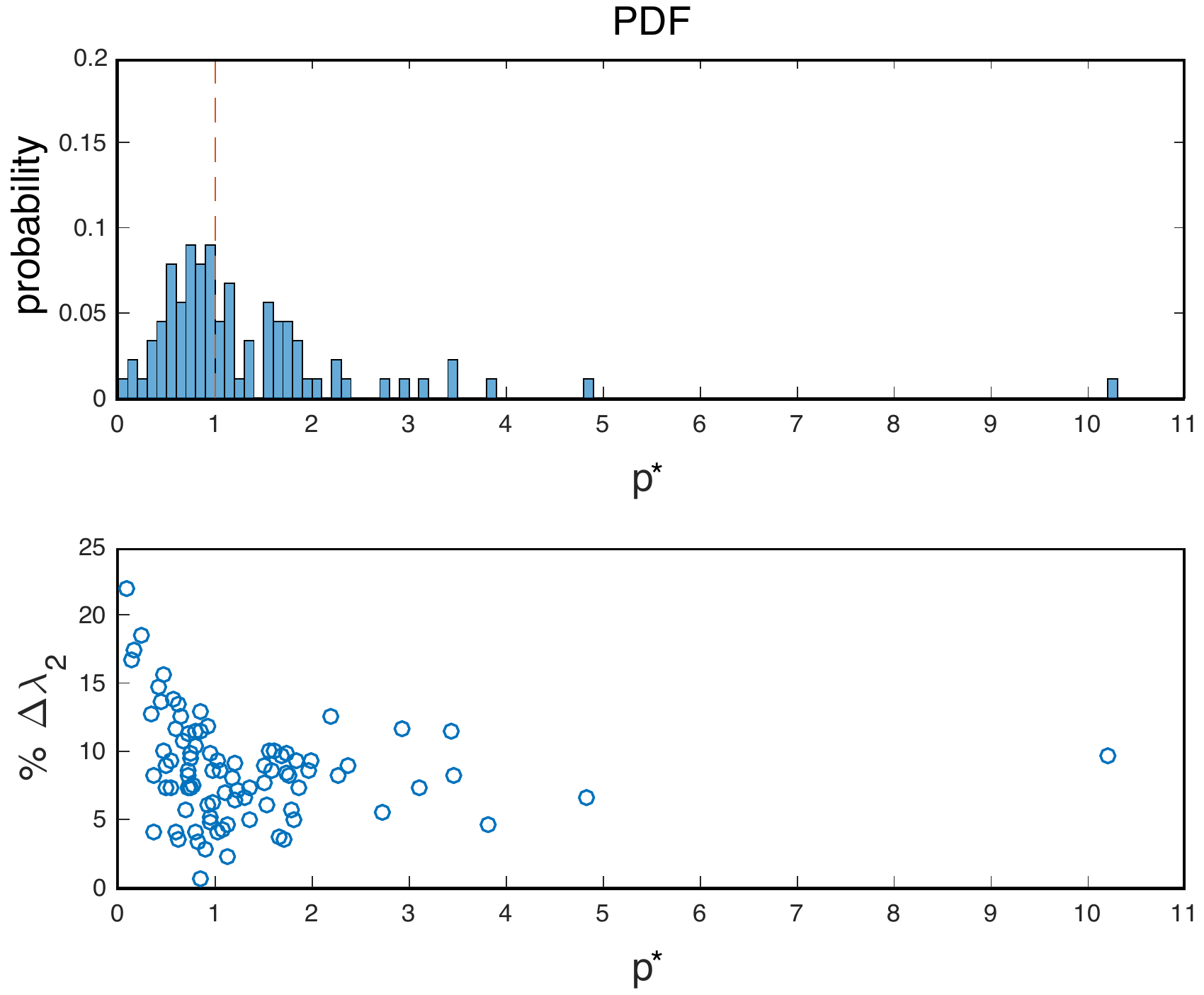}
\caption{{\bf MEG of full multilayer networks and the transition point $p^*$.}
(Top) Probability distribution function (PDF) of the transition point $p^*$ for 2-layer networks (of the alpha and beta layers) of the $89$ individuals.
The dashed line corresponds to $p^*=1$. (Bottom) Percentage of deviation of $\lambda_2$ with respect to the value of
the aggregated network: $\% \Delta \lambda = (\lambda_{2,\mathrm{\mathrm{agg}}} - \lambda_2) /\lambda_{2,\mathrm{\mathrm{agg}}}$, where $\lambda_2$ denotes the algebraic connectivity of the full multilayer network. Each circle represents one of the $89$ individuals.
}
\label{fig:fig07}
\end{figure}

\textcolor{black}{In Fig.~\ref{fig:fig07}, we show} (top panel) the PDF of the values of $p^*$ calculated for the multilayer networks of each of the $89$ individuals and (bottom panel) the percentage of deviation of $\lambda_2$ with respect to the value $\lambda_{2,\mathrm{agg}}$ of the aggregated network. 
\textcolor{black}{We observe that the peak of the PDF for $p^*$ is near $p=1$. That is, the multilayer networks that we construct from empirical data sets are close to the transition point}. \textcolor{black}{This figure also illustrates discrepancies in the values of $\lambda_2$ between the aggregated and multilayer networks, illustrating that it is necessary to differentiate between the two cases when interpreting the value of $\lambda_2$, especially when comparing results from different studies.}


\begin{table}[!h] \centering
\begin{tabular}{|c|c|c|c|c|}
\hline \hline
 	Layer	&  Intralayer strength	&  Interlayer strength		\\ 
\hline \hline
theta ($\theta$)	&	12.69\%	&	9.53\%	 \\ 
\hline
alpha ($\alpha$)	&  	31.43\%		& 	7.49\%	 \\ 
\hline
beta ($\beta$)	&  12.18\%	 	& 9.34\%  \\ 
\hline
gamma ($\gamma$)	&  	6.85\%		&	 10.47\%  \\
\hline \hline
\end{tabular}
\caption{{\bf Percentage of strength of each layer in the (4-layer) multilayer network from MEG data.} The first column indicates the layer. The second column indicates the percentage of strength of all intralayer edges in the full 4-layer multilayer network that come from that layer; that is, for a layer $l \in \{\theta, \alpha, \beta, \gamma\}$, the percentage is given by 
$100 \times \frac{\|{{\bf M}^{l} \|_1} }{\|{\bf M}^{\theta\alpha\beta\gamma}\|_1}$, where the operator ${\|{\bf M}\|_1}$ is the entrywise $1$-norm of $\bf M$, corresponding to the sum of all elements of the matrix.
The third column gives the percentage of the strength of the interlayer edges. 
That is, for a layer $l'$, the percentage is $100 \times \frac{2 \sum_{l \neq l'} \|{\bf P}^{l'l} \|_1}{ \|{\bf M}^{\theta\alpha\beta\gamma} \|_1}$, where $l' \in \{\theta, \alpha, \beta, \gamma\}$. Each percentage in the table is a mean over the 89 individuals.
}
\label{tab:tab02}
\end{table}

\begin{figure}[!t]
\centering
\includegraphics[angle=0,width=0.50\textwidth]{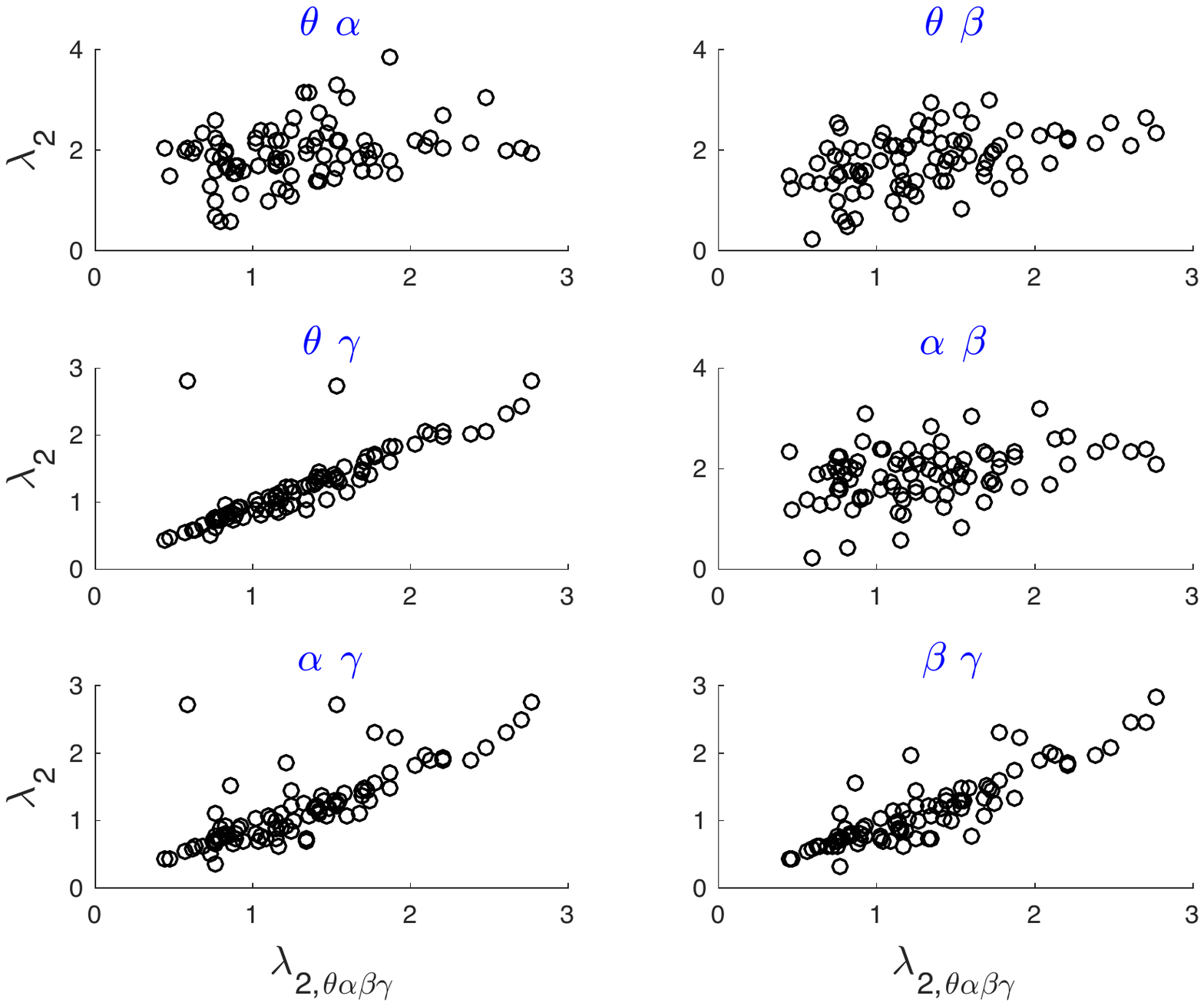}
\caption{{\bf Combining different frequency bands into multilayer networks.}
Algebraic connectivity $\lambda_2$ of all possible combinations of 2-layer networks versus $\lambda_{2,\theta\alpha\beta\gamma}$ of the 4-layer full multilayer networks. We construct the layers from the theta ($\theta$), alpha ($\alpha$), beta ($\beta$), and gamma ($\gamma$) frequency bands. The labels at the top of each plot correspond to the associated frequency bands. Each circle corresponds to one of the 89 individuals. 
}
\label{fig:fig08}
\end{figure}

Finally, we investigate how the combination of layers from different frequency bands affects the value of $\lambda_2$.
In our analysis thus far, we have focused on a 2-layer
network formed by alpha and beta frequency bands, because they are often associated to brain activity during the resting state. Nevertheless, because the signal has been filtered into
$4$ different frequency bands, there are 8 possible combinations of 2 layers. 
In Fig.~\ref{fig:fig08}, we show the relation between $\lambda_2$ for all possible combinations of 2-layer networks versus that for the full 4-layer multilayer network. Observe the strong correlation of the 2-layer networks that include the gamma layer (especially the one that consists of the theta and gamma layers) with the full multilayer network. One can explain such a correlation by inspecting
the total strength of each frequency band. In Table \ref{tab:tab02}, we separate the intralayer and interlayer strengths to facilitate interpretation of the results. We observe that the gamma band is the least-active layer, as it is the one with the lowest intralayer strength. Nevertheless, it has the highest interlayer strength (i.e., the sum of the weights of its interlayer edges with all other layers is the largest), so it is the layer that appears to interact most strongly with the other layers. 
Because (as we have seen) the full multilayer functional networks are close to the mean  $\langle p^* \rangle$ of the transition point, the weight of the intralayer
connections has a strong influence on the value of $\lambda_2$. Therefore, the 2-layer theta--gamma network, which includes the layers with the highest interlayer strengths, is the one with the strongest correlation of $\lambda_2$ with the full multilayer network [see Fig.~\ref{fig:fig08}(c)].

\section*{Conclusions and Discussion}

Using network analysis as a tool for analyzing brain-imaging data, and (more specifically) implementing and studying a multilayer description of brain activity has both advantages and drawbacks that must be investigated carefully. As we have discussed, it is possible to encode such information as either a multiplex network or as more general types of multilayer networks, but different choices lead to different results, which then must be interpreted from a neuroscientific perspective. In our paper, we have performed such an analysis to explore the implications of the two approaches on spectral information --- and specifically on the algebraic connectivity $\lambda_2$, which has been related to structural, diffusion, and synchronization properties of networks \cite{newman2010,miegh2011,masuda2017,almendral2007,radichi2013,gomez2013}. 
We have seen how the heterogeneity of the interlayer edges of multiplex networks leads 
to deviations of the theoretical predictions obtained when all interlayer edges have an equal weight $p$, and we observed that the deviation is even larger when interlayer edges are missing.
The importance of these results, which imply large differences in qualitative dynamics, is underscored by the fact that both heterogeneity and missing interlayer edges are common features of brain-imaging data.  

It is also important to understand that, \textcolor{black}{although a multiplex description of brain networks is an important and useful approach to integrate multivariate information, and it is often an extremely natural approach --- such as for integrating anatomical and functional networks, examining the temporal evolution of a functional network, and so on --- it is just an initial step towards developing increasingly complete models to better analyze the spatial and temporal complexity of brain networks \cite{papo2014}.
In this quest, a natural step, although certainly not the final step, is to use a full multilayer description for subsequent analyses of frequency-based brain networks\footnote{It is also desirable, for example, to examine this type of network in a time-resolved manner (ideally using continuous time), to incorporate spatial constraints, and so on.}. In this type of network, it is more biologically plausible to represent brain activity as a full multilayer network rather than as a multiplex network. The reasons are twofold:}
 (i) a brain region does not necessarily coordinate with 
itself at different frequencies; and (ii) there may also exist cross-frequency coupling between different brain regions. 
Consequently, interlayer connections do not necessary follow the multiplex paradigm, because some of the interlayer edges probably should not be present, whereas others should be included to account for interlayer coupling between different nodes (i.e., different brain regions), thereby necessitating a fully general multilayer approach.\textcolor{black}{\footnote{Analogous critiques are also relevant for research, including our own, on multilayer analysis of time-dependent brain networks (see, e.g., \cite{bassett2011}), for which one can envision incorporating time delays in coordination between brain regions.}} \textcolor{black}{Nevertheless, the use of frequency-based multiplex networks has been an effective strategy in previous work for successfully distinguishing between healthy individuals and individuals who are suffering from schizophrenia \cite{dedomenico2016}, so there are times when this approach works well.}

As we have illustrated in our computations, because the number of interlayer edges scales with $N$ in multiplex networks but with $N^2$ in the multilayer networks, there are important quantitative differences in the values of the algebraic connectivity in the two types of networks. Nevertheless, the effects of heterogeneity and missing interlayer edges on full multilayer networks are qualitatively similar to those in multiplex networks. Specifically, the analytical predictions that we obtained with homogeneous multilayer networks deviate from those of heterogeneous cases, showing an important discrepancy for values of the coupling-strength parameter $p$ that are larger than the transition point $p^*$. Again, the effect of missing interlayer edges is particularly dramatic for $p \geq p^*$.

Our analysis of experimental-imaging data confirms our results with synthetic networks.
The small number of interlayer connections in a multiplex network leaves the system in a region in which layers behave as if they are structurally independent, and a deeper analysis reveals that the weights of the interlayer edges from the experimental data need to be increased by several orders of magnitude to reach the transition point $p^*$. 
\textcolor{black}{Consequently, using a multiplex representation leads to a network with structurally independent layers, unless one uses extremely large weights (i.e., $p\gg1$) for the interlayer edges. However, the weights of the interlayer edges cannot be increased artificially without limits, because then they would not have a biophysical meaning when compared with the weights of the intralayer edges.}

Interestingly, \textcolor{black}{our results change drastically when using a full multilayer description, in which a brain region whose dynamics are in a given frequency band can now couple with other brain regions in a different frequency band.} When one considers all possible interlayer edges, the majority of the $89$ individuals in the experiment are close to the 
transition point $p^*$, even though the percentage of weight in the interlayer edges is always about $10$\% of that of the intralayer ones (see Table \ref{tab:tab02}). As we showed in our analysis of synthetic data, it is around this point that heterogeneity and missing interlayer edges begin to yield important discrepancies with the theoretical results from homogeneous multilayer networks [see Fig.~\ref{fig:fig04}(a)]. 

It is important to note that the percentage of total weight of interlayer edges (relative to the total weight of all edges) depends strongly on how one measures coordination 
between brain regions. 
When one uses diagnostics other than mutual information, it is possible to observe differences in this percentage, which may move multilayer networks above or below the transition point $p^*$. Thus, no matter what measure one uses, it is mandatory to first analyze the percentage of total weight of the interlayer edges and to interpret the value of $\lambda_2$ with respect to the value of $p^*$.

The fact that, in frequency-based multilayer networks, the weight of interlayer edges strongly influences the value of $\lambda_2$ despite being much smaller than the weights of intralayer edges (see Table \ref{tab:tab02}) highlights the importance of adequately evaluating cross-frequency coupling, which has traditionally been disregarded when representing brain activity in terms functional networks. 
As has been discussed prominently in neuroscience (including in critiques of connectomics) \cite{kopell2014}, the dynamics matter, and investigate of dynamics must include incorporation and analysis of coupling between different frequency bands.
The methodology that is used to quantify interactions between brain regions at different frequencies leads to different values of $\lambda_2$; and there are also other important dynamical issues, such as phase--amplitude correlations, that we have not investigated in this paper. (See \cite{bastos2016} for a review of how common reference, volume conduction, field spread, or common input affect the quantification of coordinated activity between brain regions.) These caveats notwithstanding, our analysis illustrates an approach for examining the effects of such phenomena on $\lambda_2$ (and hence on spectral structure) in multilayer functional networks, and we expect that similar qualitative phenomena will arise in both multiplex and full multilayer networks constructed using other choices (e.g., different measures than MI) from the ones that we chose in order to provide a concrete illustration.

We initially studied 2-layer networks of the alpha and beta frequency bands, because it is known that these bands incorporate a large amount of the power spectrum of brain activity during resting state and these frequency bands are the ones that typically exhibit stronger synchronization between brain regions (e.g., see \cite{brookes2016}) than the others. The alpha and beta bands are thus the most commonly-studied frequency bands in resting-state studies, and we followed this tradition.
Nevertheless, in our particular case, our comparison between 2-layer and 4-layer networks illustrates that the gamma band is the one that most influences the spectral properties of the full 4-layer multilayer network (see Fig.~\ref{fig:fig08}). 
As we showed in Table \ref{tab:tab02}, the theta and gamma layers are the ones whose interlayer edges 
give the largest contribution,
thereby leading to the strongest correlation between a 2-layer network (with theta and gamma layers) and the complete (4-layer) multilayer network.
This fact highlights the importance of the well-known phase--amplitude correlations between the theta and gamma frequency bands \cite{canolty2006,aru2015}, as the former acts as a carrier of fast-amplitude fluctuations in the latter. Consequently, theta--gamma coupling may be fundamental for understanding the multilayer nature of functional brain networks.

\textcolor{black}{It also worth mentioning that the facts that we are constructing frequency-based multilayer networks from (i) data at the sensor level and (ii) using the MI as a measure to quantify mutual interdependency between brain regions lead to unavoidable errors in the quantification of edge weights due to signal mixing and spurious correlations 
from common sources. However, to date, there does not exist an error-free methodology to construct functional brain networks and using other alternatives such as source reconstruction or different measures to evaluate interdependency between brain regions have their own drawbacks. (for a detailed discussion, see ``Methodological Considerations'' in Materials and Methods.)}
 
Finally, although our calculations with experimental data used resting-state MEG recordings, we expect to observe similar behavior in frequency-based multilayer networks of different origins --- whether obtained from any of a large variety of different cognitive or motor tasks, with different brain-imaging techniques, or even if they come from a completely different system (such as functional climate networks \cite{tsonis2004}). \textcolor{black}{ However, despite the generality of our results, it is important to examine richer models of frequency-based brain network, such as ones that include spatial constraints and temporal evolution.}

\section*{MATERIALS AND METHODS}

{\bf{Data Acquisition.}} The data sets were made available by the Human Connectome Project (HCP); see \cite{hcp} and \cite{larson2013} for details. 
The experimental data sets consist of magnetoencephalographic (MEG) recordings of a group of $89$ individuals, during resting state, for a period of approximately $2$ minutes. During the scan, subjects were supine and maintained fixation on a projected red crosshair on a dark background. Brain activity was scanned on a whole-head MAGNES 3600 (4D Neuroimaging, San Diego, CA, USA) system housed in a magnetically-shielded room, and it included up to 248 magnetometer channels. The root-mean-squared (RMS) noise of the magnetometers is about 5 fT/sqrt (Hz) on average in the white-noise range (above 2 Hz). The data were recorded at sampling rate of $f_s \approx 508.63$ Hz. Five current coils attached to a subject, in combination with structural-imaging data and head-surface tracings, were used to localize the brain in geometric relation to the magnetometers and to monitor and partially correct for head movement during MEG acquisition. Artifacts, bad channels, and bad segments were identified and removed from the MEG recordings, which were processed with a pipeline based on independent component analysis (ICA) to identify and clean environmental and subject artifacts \cite{larson2013}. \textcolor{black}{After this process, the number $N$ of channels considered for each individual was in the range \textcolor{black}{$235 \leq N \leq 246$} (with a mean of 243.42), because some of them were used as references and others were disregarded.} 

\vspace{.1 in}

{\bf{Coordination between brain regions.}} To estimate coordination between brain regions, we first apply a band-pass filter to the preprocessed signals to obtain, for each of the $N$ sensors, a set of four different time series, each of which corresponds to a specific frequency band: theta [3--8] Hz, alpha [8--12] Hz, beta [12--30] Hz, and gamma [30--100] Hz. 
We thereby obtain $4N$ time series of $t=149646$ points for each of the $q=89$ individuals.
We then order the filtered signals according to their corresponding frequency band, so the time series $X^s$ with $s \in \{1, \dots ,N\}$ corresponds to the theta band, $X^s$ with $s \in \{N+1,\dots,2N\}$ corresponds to the alpha band, $X^s$ with $s \in \{2N+1,\dots,3N\}$ corresponds to the beta band, and $X^s$ with $s \in \{3N+1,\dots,4N\}$ corresponds to the gamma band. Each of the $4N$ time series corresponds to a node of the frequency-based networks.
We calculate the mutual information (MI) between the time series $X^i$ and $X^j$ of a pair of nodes $i$ and $j$ with the formula
\begin{equation} \label{E:RFV}
	\mathrm{MI}_{ij} = \sum_{uv} p_{uv} \log\left(\frac{p_{uv}}{p_u p_v}\right)\,, 
\end{equation}
where $p_{u}$ is the probability that
$X^i=x_u$, the quantity $p_{v}$ is the probability that
$X^j=x_v$, and $p_{uv}$ is the joint probability that
$X^i=x_u$ and $X^j=x_v$ at the same time step.
We set the number of bins of the
probability distribution functions to be $u=v=5 \times \lfloor (\sqrt{t/10}) \rceil$, where $t$ the is the number of time steps and $\lfloor y \rceil$ corresponds to the nearest integer function of the real number $y$ (where we round up for $.5$).
When $X^i$ and $X^j$ are independent variables, $p_{uv}=p_u p_v$, and the resulting $\mathrm{MI}_{ij}$ is $0$. When $X^i=X^j$ for each time series, the $\mathrm{MI}_{ij}$ achieves its maximum value. Note that $\mathrm{MI}_{ij}=\mathrm{MI}_{ji}$, so we obtain an undirected edge between the two time series, and we are disregarding causality. Calculating $MI_{ij}$ allows one to detect coordinated
activity even for time series that include different frequency bands. See \cite{pereda2005,bastos2016} for a review of different measures for quantifying coordination between brain regions and a discussion of their advantages and pitfalls.
Note that $\mathrm{MI}_{i(i+N)}$ measures the coordination between two different
frequency bands in the same brain region $i$.
\textcolor{black}{After calculating $\mathrm{MI}_{ij}$ using Eq.~\eqref{E:RFV}, we \textcolor{black}{generate surrogates (which we subsequently as a threshold for mutual information)} using a block-permutation procedure \cite{canolty2006}: we simultaneously cut each time series into blocks of $1018$ points (about $2$ seconds), and we permute the resulting blocks uniformly at random. This procedure preserves lower frequencies and time-series features (such as nonstationarity and nonlinearity) below the chosen temporal scale.
(See the Supplementary Information for further discussion and a detailed investigation of the influence of block length on the surrogates.)
We then evaluate the \textcolor{black}{mutual information} between each surrogate time series to obtain $\mathrm{MI}^{\mathrm{rand}}$.}

\vspace{.1 in}

{\bf{Frequency-based multilayer networks.}} We construct a frequency-based multilayer network for each individual from the matrix with the MI of each pair of sensors for the four different frequencies bands. 
Each layer includes nodes with the same frequency band, yielding four different layers: theta ($\theta$), alpha ($\alpha$), beta ($\beta$), and gamma ($\gamma$). 
\textcolor{black}{We now use $\mathrm{MI}^{\mathrm{rand}}$ of the surrogate times series as a threshold and} construct a weighted supra-adjacency matrix {\bf W} with elements ${\bf W}_{ij}=\mathrm{MI}_{ij} - \mathrm{MI}^{\mathrm{\mathrm{rand}}}_{ij}$ if $\mathrm{MI}_{ij} > \mathrm{MI}^\mathrm{\mathrm{rand}}_{ij}$ and $0$ otherwise. We thereby account only for edges with statistically-significant edges. (In Fig.~\ref{fig:fig02}, one can see what fractions of edges are $0$ in each case.) Finally, we apply a linear normalization to {\bf W} to obtain
\begin{equation} \label{eq:normalization}
	{\bf M}_{ij} = \frac{{\bf W}_{ij} - w_{\mathrm{min}}}{w_{\mathrm{max}}-w_{\mathrm{min}}}\,, \nonumber
\end{equation}
where $w_{\mathrm{min}}$ and $w_{\mathrm{max}}$ are, respectively, the largest and smallest values of {\bf W}. This ensures that
${\bf M}_{ij} \in [0,1]$ for each individual, which facilitating comparisons between them.
The weighted supra-adjacency matrix ${\bf M}$ includes some number of $0$ entries, which account for interactions between brain regions and frequencies that we deem to not be statistically significant. It also has four blocks along the diagonal that encode interactions within each layer (i.e., the same frequency band for different brain regions), and it has off-diagonal blocks
that quantify coordination between different frequencies. See Fig.~\ref{fig:fig01}(b) for a schematic.

{\bf{Methodological considerations.}} \textcolor{black}{An important issue is the applicability of our results to functional brain networks constructed using different approaches, such as by using source reconstruction or with different ways of evaluating coordination between brain regions. \textcolor{black}{In our study, we constructed functional brain networks from magnetoencephalographic recordings measured at the sensor level, instead of using the actual magnetic field generated at each brain region (i.e., source), which can be inferred using source-reconstruction methods.} While working with time series at the sensor level necessarily entails signal mixing \cite{schoffelen2009}, the use of source reconstruction has what is known as the ``inverse problem" (of inferring the actual magnetic field created by the brain regions), and it thus requires the introduction of a priori assumptions in the \textcolor{black}{model used for the source reconstruction}.} 
\textcolor{black}{This issue has led to a diversity of algorithms \textcolor{black}{to obtain source-reconstructed time series and, in many cases, different algorithms yield qualitatively different time series at the source level}. See, for example, \cite{schoffelen2009} and \cite{belardinelli2012}.}
\textcolor{black}{Consequently, there is an open debate about what constitutes the most appropriate methodology to construct functional networks using source reconstruction \cite{palva2012,vandiessen2015}. 
However, regardless of how one constructs a functional brain network (and whether one uses sensors or sources), our analysis has the same qualitative implications.
Specifically, estimating the value of $\lambda_2$ from matrices associated with function brain networks is affected strongly by edge-weight heterogeneity and missing interlayer edges. One also observes the same qualitative differences between a multiplex construction and a full multilayer one, as this phenomenon depends on the number of interlayer edges and not on specific details of the construction of functional brain networks.}
\textcolor{black}{
A similar reasoning applies to the construction of frequency-based brain networks using different ways of quantifying coordination between brain regions. We used MI because it is able to successfully capture both linear and nonlinear correlations, it is \textcolor{black}{an adequate method for quantifying the interdependency of signals that are split into frequency bands} \cite{david2004}, and it has been used widely in the past to construct functional brain networks. 
(See, e.g., \cite{chai2009,deuker2009,basset2009,jin2011,becker2012}.) 
On the down side, MI cannot avoid zero-lag correlations that originate from common sources. (This flaw also occurs in many other commonly-employed similarity measures, such as partial correlation, coherence, and phase-locking value \cite{bastos2015}.) 
Although the systematic deletion of zero-lag correlations is a very strict assumption, because not all zero-lag correlations are due to common sources or volume conduction \cite{vicente2008,christodoulakis2013,porz2014}, there does not exist a consensus on how
to eradicate them (see, e.g., \cite{stam2007,brookes2012}). 
It is thus important for future work to conduct a systematic comparison between the most widespread coordination measures and their consequences on the spectral properties of functional networks.
As we stated previously, the effect that edge-weight heterogeneity and missing interlayer edges have on functional multiplex and multilayer networks is a general phenomenon, as it does not rely on the specific similarity measure used to evaluate coordination between brain regions, as long as there is sufficient heterogeneity in the weights of interlayer edges. As \textcolor{black}{shown in \cite{brookes2016}, such heterogeneity is expected} in frequency-based functional brain networks, given the complicated interactions between different brain regions at different frequencies.}


\section*{ACKNOWLEDGEMENTS}

We thank John Allen for fruitful conversations. J.M.B. acknowledges financial support from Spanish MINECO (project FIS2013-41057) and from Salvador de Madariaga Program (PRX15/00107), which allowed him to visit University of Oxford in summer 2016. We also thank the referees for their helpful suggestions.


\end{document}